\documentclass[times,final]{elsarticle}
\usepackage{framed,multirow}
\usepackage{amsmath}
\usepackage{subfigure}
\usepackage{amssymb}
\usepackage{latexsym}
\usepackage{epstopdf}

\usepackage{url}
\usepackage{xcolor}
\definecolor{newcolor}{rgb}{.8,.349,.1}

\newcommand{\secref}[1]{\S\ref{#1}}

\begin{document}

%\verso{Given-name Surname \textit{etal}}

\begin{frontmatter}

\title{A QMC-deep learning method for diffusivity estimation in random domains}%
\author[1,2]{Liyao Lyu}
\ead{lylv@stu.suda.edu.cn}
\cortext[cor1]{Corresponding authors}
\author[3]{Zhiwen Zhang \corref{cor1}}
\ead{zhangzw@hku.hk}
\author[1,4]{Jingrun Chen \corref{cor1}}
\ead{jingrunchen@suda.edu.cn}

\address[1]{School of Mathematical Sciences, Soochow University, Suzhou, China}
\address[2]{CW Chu College, Soochow University, Suzhou, China}
\address[3]{Department of Mathematics, The University of Hong Kong, Pokfulam Road, Hong Kong SAR, China}
\address[4]{Mathematical Center for Interdisciplinary Research, Soochow University, Suzhou, China}

%\received{1 May 2013}
%\finalform{10 May 2013}
%\accepted{13 May 2013}
%\availableonline{15 May 2013}
%\communicated{S. Sarkar}

\begin{abstract}
%%%
Exciton diffusion plays a vital role in the function of many organic semiconducting opto-electronic devices, where an accurate description requires precise control of heterojunctions. This poses a challenging problem because the parameterization of heterojunctions in high-dimensional random space is far beyond the capability of classical simulation tools. Here, we develop a novel method based on quasi-Monte Carlo sampling to generate the training data set and deep neural network to extract a function for exciton diffusion length on surface roughness with high accuracy and unprecedented efficiency, yielding an abundance of information over the entire parameter space. Our method provides a new strategy to analyze the impact of interfacial ordering on exciton diffusion and is expected to assist experimental design with tailored opto-electronic functionalities.
%%%%
\end{abstract}

%\begin{keyword}
%% Keywords
%\KWD Keyword1\sep Keyword2\sep Keyword3
%\end{keyword}

\end{frontmatter}

%\linenumbers

\section{Introduction}

Over the past decades, much attention has been paid on organic semiconductors for applications in various opto-electronic devices~\cite{Jason2012,Su2012,Martin1999,Forrest2004911}. These materials include small molecules~\cite{Peumans20033693, Lin2014280}, oligomers~\cite{Sanaur2006,Murphy2007}, and polymers~\cite{Pettersson1999487, Mikhnenko20126960}. Exciton diffusion is one of the key processes behind the operation of organic opto-electronic devices~\cite{DeOliveiraNeto20123039,Brabec200550,Bredas20091691}. From a microscopic perspective, exciton, a bound electron-hole pair, is the elementary excitation in opto-electronic devices such as light emitting diodes and organic solar cells. The exciton diffusion length (EDL) is the characteristic distance that excitons are able to travel during their lifetime~\cite{Lin2014280}. A short diffusion length in organic photovoltaics limits the dissociation of excitons into free charge~\cite{Terao2007,Menke2013152}. Conversely, a large diffusion length in organic light emitting diodes may limit luminous efficiency if excitons diffuse to non-radiative quenching sites~\cite{Enhancedcarrier}.
\par
As quasi-particles with no net charge, excitons are difficult to probe directly by electrical means~\cite{Mullenbach2017}. This is particularly true in organic semiconductors where the exciton binding energy is $\sim$1 electronvolt~\cite{Tang1986183}. Reported techniques to measure EDL include photoluminescence (PL) surface quenching~\cite{Terao2007,Rim2007,Markov2005,Scully2006,Goh2007,Shaw20083516,Wu2005,Theander200012957,Lin2014280}, time-resolved PL bulk quenching modeled with a Monte Carlo (MC) simulation~\cite{Mikhnenko20126960,Mikhnenko201214196}, exciton-exciton annihilation~\cite{Cook200933,Lewis2006452,Masri20131445,Shaw2010155},  modeling of solar cell photocurrent spectrum~\cite{Qin20111967,Stubinger20013632,Yang20031737,Halls19963120,Theander200012957,Ghosh19785982,Wagner1993423,Bulovic199588,Pettersson1999487,Peumans20033693,Rim2008,Huijser2008}, time-resolved microwave conductance~\cite{Fravventura20122367,Kroeze20037696}, spectrally resolved PL quenching~\cite{Lunt20101233,Bergemann2011,Rand20122987} and F\"{o}rster resonance energy transfer theory~\cite{Lunt2009,Lunt20101233,Mullenbach20133689}.
From a theoretical perspective, the minimal modeling error is given by the diffusion equation model~\cite{Chen2016754}, which is employed in the current work.
\par
To be precise, the device used in PL surface quenching experiment includes two layers of organic materials with thickness ranging from dozens of nanometers to hundreds of nanometers. One layer of material is called donor and the other is called acceptor or quencher according to the difference of their chemical properties. Under the illumination of solar lights, excitons are generated in the donor layer and diffuse in the donor. Due to the exciton-environment interaction, some excitions die out and emit photons which lead to the PL. The donor-acceptor interface serves as the absorbing boundary while other boundaries serve as reflecting boundaries due to the tailored properties. Since the donor-acceptor interface is not exposed to the air/vacuum and the resolution of the surface morphology is limited by the resolution of atomic force microscopy, the interface is subject to an uncertainty. It is found that the fitted EDL is sensitive to the uncertainty in some scenarios. From a numerical perspective, the random interface requires a parametrization in high-dimensional random space, which is prohibitively expensive for any simulation tool. For example, MC method overcomes the curse of dimensionality but has very low accuracy~\cite{HONG2014}. Stochastic collection method has high accuracy but is only affordable in low dimensional random space~\cite{SC3}. Asymptotics-based method is efficient but its accuracy relies heavily on the magnitude of randomness~\cite{Chen2019894}. In the current work, we
propose a novel method based on deep learning with high accuracy and unprecedented efficiency.
\par
Recently, increasing attentions have been paid to apply machine learning (ML) techniques to materials-related problems. For example, the classification of crystal structures of transition metal phosphide via support vector machine~\cite{Oliynyk201717870} leads to the discovery of a novel phase~\cite{Oliynyk20166672}. Likewise, a hybrid probabilistic model based on high-throughput first-principle computation and ML was developed to identify stable novel compositions and their crystal structures~\cite{Hautier20103762}. Physical parameters such as band gap~\cite{Isayev2017,Xie2018}, elastic constants~\cite{Isayev2017,DeJong2016}, and Debye temperature~\cite{Isayev2017} have also been predicted using an array of ML techniques. In another line, deep learning (DL) in computer science has had great success in text classification~\cite{Wang2012EndtoendTR}, computer vision~\cite{NIPS2012_4824}, natural language processing~\cite{Sarikaya:2014:ADB:2687012.2687014}, and other data-driven applications \cite{Cosmin2019,Xiaoyingzhuang2019}. Recently, DNNs have been applied to the field of numerical analysis and scientific computing, like inverse problems \cite{LiangYan} and dimensional partial differential equations (PDEs)\cite{Carlos2019,Han2018,Keli2020,WANG2020108963}. One significant advantage of DL is its strong ability to approximate a complex function in high dimensions and extract features with high precision using composition of simple nonlinear units. Meanwhile, benefiting from recent advances in parallel graphics processing unit - accelerated computing, huge volumes of data can be put into the DL architecture for training.
\par
In this work, we employ DL to extract a complex function of EDL in terms of the random interface parametrized in a high-dimensional space.
The fitted function has rich information, which explains a few interesting experimental observations. Compared to classical simulation tools, our approach
has the following features: quasi-Monte Carlo (QMC) sampling~\cite{niederreiter1992random, Russel1998, dick2013high} for data collection and ResNet~\cite{He2015} for training. The size of data in the former step grows only linearly with respect to the dimension of random space, thus our approach overcomes the curse of dimensionality (with possibly a logarithmic growing factor depending on the dimension). With the usage of ResNet in the latter step, a complex function can be extracted with high accuracy.

Our main contribution is the use of QMC sampling to explore the high-dimensional space of surface roughness, i.e., generating realizations of surface roughness by the QMC method; see the left column in Fig. \ref{fig:schematic}. Once a surface roughness is generated, an inverse problem is solved to produce EDL. DL is then used to construct a map between surface roughness and EDL; see the center column in Fig. \ref{fig:schematic}. Afterwards, the full landscape of EDL on surface roughness can be generated and further analyzed; see the right column in Fig. \ref{fig:schematic}. Generally speaking, QMC method has an approximation accuracy independent of dimension which scales inversely with respect to the number of samples (with possibly a logarithmic factor depending on dimension). This significantly outperforms Monte-Carlo method. Other sampling strategies, such as Latin Hypercube Sampling, Hammersley Sequence Sampling, and Latin hypercube-Hammersley sequence sampling, shall also perform better. ResNet is used to construct the map between surface roughness and EDL. Note that the data generation dominates the whole simulation since each data requires solves an inverse problem involving solving an elliptic problem over a curved domain. Therefore, even though there may be better network structures in the sense of accuracy or efficiency, we stress that the one used in this paper has already provided a great choice and only linear growth of number of samples with respect to dimension is needed for accurate training. No significant change is found if more layers or parameters are added in the ResNet structure we have used.

The rest of this paper is organized as follows. In \secref{Sec:Model Description}, we introduce models for exciton diffusion in 1D, 2D, and 3D, respectively.
In \secref{Sec:Method}, we propose a QMC-machine learning method based on QMC sampling and ResNet. Using a series of numerical simulations in \secref{Sec:Result and Discussion}, we show that the QMC-machine learning method is both accurate and efficient. Moreover, modeling error and mode dependence can be extracted
from the fitted EDL, which explains a few interesting experimental observations. Conclusions are drawn in \secref{Sec:Conclusion}.

%--------------------------------------------------------------------------

\section{Model description}
\label{Sec:Model Description}
\subsection{1D model}
An exciton that diffuse in the donor follows a diffusion-type equation over a 1D random domain
\begin{equation}\label{eqn:1DInterface}
  D_1=\{x:x\in[\theta(\omega), d]\},
\end{equation}
where the interface is reduced to a random point $x=\theta(\omega)$.
\par
The corresponding diffusion equation is
\begin{equation}
\left\{
\begin{aligned}\label{eqn:one dimensional model}
&\sigma^2u_{xx} -u + G(x) =0 , & \theta(\omega)<x<d,\\
&u_x(d)=0 ,\\
&u(\theta(\omega))=0 ,
\end{aligned}\right.
\end{equation}
where $\sigma$ is the EDL to be extracted, $u$ is the exciton density, $G$ is the normalized exciton generation function by the transfer matrix method~\cite{burkhard_accounting_2010}. $x=d$ serves as the reflecting boundary and Neumann boundary condition is imposed on the boundary exposed in air. $x=\theta(\omega)$ serves as the absorbing boundary and homogenous Dirichlet boundary condition is imposed on the donor-acceptor interface and $\theta$ is the input of the network in the 1D case. The photoluminescence (PL) is
\begin{equation}
\begin{aligned}\label{eqn:1Dphotoluminescence}
I_{\theta(\omega)}[\sigma,d]= \int_{\theta(\omega)}^{d} u(x) \mathrm{d}x,
\end{aligned}
\end{equation}
where the integral is approximated by the Simpson's rule in the simulation.
The 1D model is commonly used to extract the exciton diffusion length (EDL) due to its simplicity and model accuracy~\cite{Lin2014280, Chen2016754}. However, the 1D model does not always work well~\cite{Lin2014280, Chen2019894} and we introduce 2D and 3D models for the same problem. They are more realistic models for exciton diffusion. We shall demonstrate their differences in \secref{Sec:Result and Discussion}.
\par
\subsection{2D model}
The 2D model is defined over a random domain
\begin{equation}\label{eqn:2DDomian}
D_2=\{(x,y)\left|h(y,\omega)<x<d,0<y<L_y\right.\},
\end{equation}
where the interface is a random line parametrized by
\begin{equation}\label{eqn:2DInterface}
h(y,\omega)=\hat{h} \sum_{k=1}^{K} k^{\beta}  \theta_{k}(\omega) \phi_{k}(y),
\end{equation}
where $\hat{h}$ is the magnitude of length due to the roughness limited by the resolution of atomic force microscopy, {$\theta_{k}(\omega)$} are i.i.d. random variables, $\phi_{k}(y)=\sin(2k\pi \frac{y}{L_y})$ and $\beta<0$ controls the decay rate of spatial modes $\phi_{k}(y)$. {Note that $\theta_k(\omega)$ are the input of the network in the 2D case}. The rougher the interface is, the closer the $\beta$ approaches $0$. Note that the dimensionless magnitude of randomness/perturbation is defined as $\varepsilon = \frac{\hat{h}}{d}$, which is used in asymptotic-based approaches \cite{Chen2020, Chen2019894}. Given the surface roughness measured in an experiment, parameters in \eqref{eqn:2DInterface} can be extracted via discrete Fourier transform.
\par

The corresponding diffusion equation is
\begin{equation}
\left\{
\begin{aligned}\label{eqn:2Ddiffusionequation}
  &\sigma^2\triangle u -u + G(x,y) =0 ,& x\in D_2,\\
  &u_x(d)=0,  \ \ u(h(y,\omega),y)=0, & 0<y<L_y,\\
  &u(x,y)=u(x,y+L_y), & h(y,\omega)<x<d,
\end{aligned}\right.
\end{equation}
and the PL is
\begin{equation}
\begin{aligned}\label{eqn:2Dphotoluminescence}
I_{\theta(\omega)}[\sigma,d]={\frac{1}{L_y}} \int_{0}^{L_y} \int_{h(y,\omega)}^{d} u(x,y) \mathrm{d}x\mathrm{d}y.
\end{aligned}
\end{equation}
Note that $\frac{1}{L_y}$ in \eqref{eqn:2Dphotoluminescence} is used for the consideration of modeling error between 1D and 2D models \cite{Chen2019894}.
\par
\subsection{3D model}
The 3D random domain can be defined as
\begin{equation}\label{eqn:3DDomian}
D_3=\{(x,y,z)\left|h(y,z,\omega_1,\omega_2)<x< d,0<y<L_y,0<z<L_z\right.\}.
\end{equation}
Here the donor-acceptor interface $x=h(y,z,\omega_1,\omega_2)$ is parameterized by
\begin{equation}
\begin{aligned}\label{eqn:3DInterface}
h(y,z,\omega_1,\omega_2)=\hat{h}\sum_{k_1=1}^{K_1} \sum_{k_2=1}^{K_2} k_1^{\beta} k_2^{\beta} \theta_{k_1}(\omega_{1})\theta_{k_2} (\omega_{2})\phi_{k_1}(y)\phi_{k_2}(z),
\end{aligned}
\end{equation}
where $\hat{h}$ is the magnitude of length due to the roughness limited by the resolution of atomic force microscopy, {$\theta_{k_1}(\omega_1),\theta_{k_2}(\omega_2)$} are i.i.d. random variables, $\phi_{k_1}(y)=\sin(2k_1\pi \frac{y}{L_y})$, $\phi_{k_2}(z)=\sin(2k_2\pi \frac{z}{L_z})$, and $\beta<0$ controls the decay rate of spatial modes $\phi_{k_1}(y), \phi_{k_2}(z)$. The rougher the interface is, the closer the $\beta$ approaches $0$. {Note that $\theta_{k_1}(\omega_1),\theta_{k_2}(\omega_2)$ are the input of the network in the 3D case.} Given the surface roughness measured in an experiment, parameters in \eqref{eqn:3DInterface} can be extracted via discrete Fourier transform.

The 3D diffusion equation reads as
\begin{equation}
\left\{
\begin{aligned}\label{eqn:model}
  &\sigma^2\triangle u -u + G(x,y,z) =0 & x\in D,\\
  &u_x(d,y,z)=0 &0<y<L_y,0<z<L_z,\\
  &u(h(y,z,\omega_1,\omega_2),y,z)=0 &0<y<L_y,0<z<L_z,\\
  &u(x,y,z)=u(x,y+L_y,z)=u(x,y,z+L_z)& h(y,z,\omega_1,\omega_2)<x<d,
\end{aligned}\right.
\end{equation}

The PL is computed by
\begin{equation}
\begin{aligned}\label{eqn:3Dphotoluminescence}
I_{\theta(\omega_1),\theta(\omega_2)}[\sigma,d]=\frac{1}{L_z}\frac{1}{L_y}\int_{0}^{L_z}\int_{0}^{L_y} \int_{h(y,z,\omega_1,\omega_2)}^{d} u(x,,y,z) \mathrm{d}x \mathrm{d}y \mathrm{d}z.
\end{aligned}
\end{equation}

At the formal level, when $L_z \rightarrow 0$, the PL of 3D model defined by \eqref{eqn:3Dphotoluminescence} reduces to the PL of 2D model defined by \eqref{eqn:2Dphotoluminescence}, and further they reduce to the PL of 1D model defined by \eqref{eqn:1Dphotoluminescence} as $L_y\rightarrow 0$.
Again, Note that $\frac{1}{L_z}\frac{1}{L_y}$ in \eqref{eqn:3Dphotoluminescence} is used for the consideration of modeling error between 1D and 3D models.

In the experiment, PL data $\{\hat{I}_i\}_{i=1}^N$ are measured by a series of bilayer devices with different thicknesses $\{d_i\}_{i=1}^N$, where $d_i$ is the thickness of the $i$-th donor layer. {Since there are not enough experimental data, we generate the reference PL data by solving Equation \eqref{eqn:one dimensional model} with a prescribed $\sigma$ and without surface roughness.}
\par
\subsection{Newton's method for the inverse problem}
The optimal EDL $\sigma$ is expected to reproduce the experimental date $\{d_i,\hat{I}_i\}_{i=1}^N$ in the sense of minimized mean square error (MSE)
\begin{equation}
\begin{aligned}\label{eqn:Inverseproblem}
\min_{\sigma} J_{\theta(\omega_1),\theta(\omega_2)}(\sigma) =\frac{1}{N} \sum_{i=1}^{N}\left(I_{\theta(\omega_1),\theta(\omega_2)}(\sigma,d_i)-\hat{I}_i\right)^2,
\end{aligned}
\end{equation}
where notations are based on 3D model.

Newton's method is used to solve \eqref{eqn:Inverseproblem} for $\sigma$.
Given $\sigma^{(0)}$, for $n=1,2,\cdots$, until convergence, Newton's method for \eqref{eqn:Inverseproblem} solves
\begin{equation*}
\begin{aligned}\label{eqn:Newtonmethod}
\sigma^{(n)} = \sigma^{(n-1)} -\alpha_n \frac{\frac{\partial}{\partial \sigma} J(\sigma^{(n-1)})}{
	\frac{\partial^2}{\partial^2 \sigma}J(\sigma^{(n-1)})}
\end{aligned}
\end{equation*}
with $\alpha_n\in(0,1]$ given by line search~\cite{Nocedal1999}.
\par

The calculated $\sigma$ is defined as $\sigma_{\theta(\omega_1),\theta(\omega_2)}$. Therefore, for different parameters $\theta(\omega_1),\theta(\omega_2)$, we get a data set $\left\{\left(\theta(\omega_1)[j],\theta(\omega_2)[j] ,\sigma_{\theta(\omega_1)[j],\theta(\omega_2)[j]}\right)_{j=1}^M\right\}$ with $M$ the size of data set.

%-----------------------------------------------------------------------------

\section{A QMC-machine learning method}
\label{Sec:Method}
\begin{figure}[ht]
\centering
\includegraphics[width=1.0\textwidth]{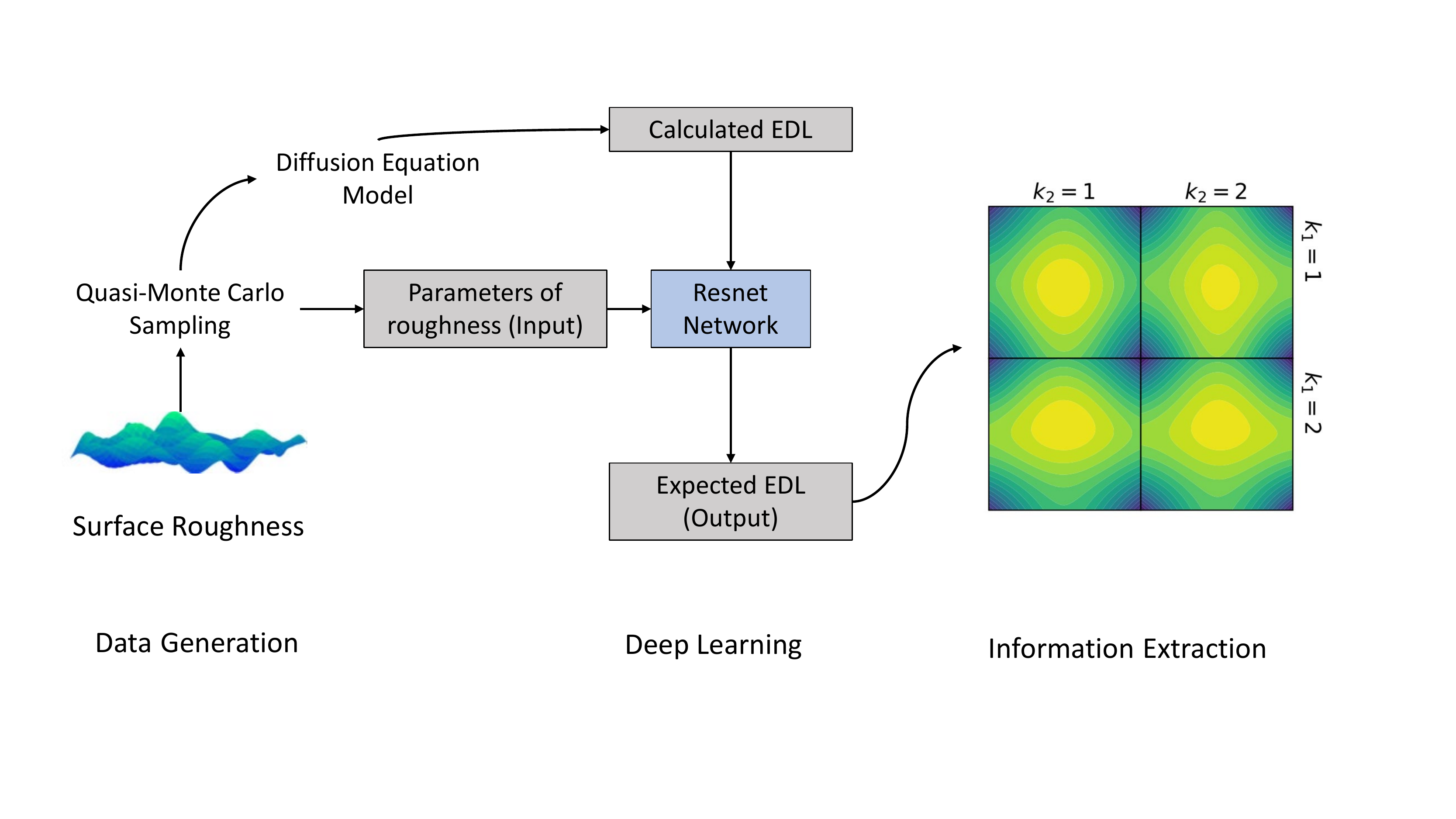}
\caption{Flow chart of the deep learning method for extracting exciton diffusion length over the parameter space.
Left: data generation; Middle: data training; Right: data prediction. In the stage of data generation, quasi-Monte Carlo method is used to sample the random space, and the actual exciton diffusion length is generated by solving the diffusion equation model. In the stage of data training, a complex function $\sigma(\theta(\omega_1),\theta(\omega_2))$ is approximated over the entire parameter space. In the stage of data prediction, given the full landscape of $\sigma(\theta(\omega_1),\theta(\omega_2))$ , both qualitative and quantitative information can be extracted.}\label{fig:schematic}
\end{figure}

The main difficulty of 2D and 3D models in random domains is the high dimension of random variables, and thus it is very difficult to solve these models with accurate results using classical simulation tools, such as MC method, stochastic collocation method, and asymptotics-based methods. Therefore, we
propose a QMC-machine learning method to overcome the curse of dimensionality.
Our approach consists of four major components: QMC sampling over the high-dimensional random space; diffusion equation model for data generation; ResNet for training to approximate a complex function of EDL; Information extraction for analysis. The flow chart of this process can be seen in Fig. \ref{fig:schematic}. The models are introduced in \secref{Sec:Model Description} and information extraction for analysis will be discussed in \secref{Sec:Result and Discussion}. For completeness, we introduce QMC sampling and ResNet in this section.

\subsection{Quasi-Monte Carlo sampling}
\begin{figure}[ht]
\centering
\subfigure[Uniform sampling]{
\includegraphics[width=0.45\textwidth]{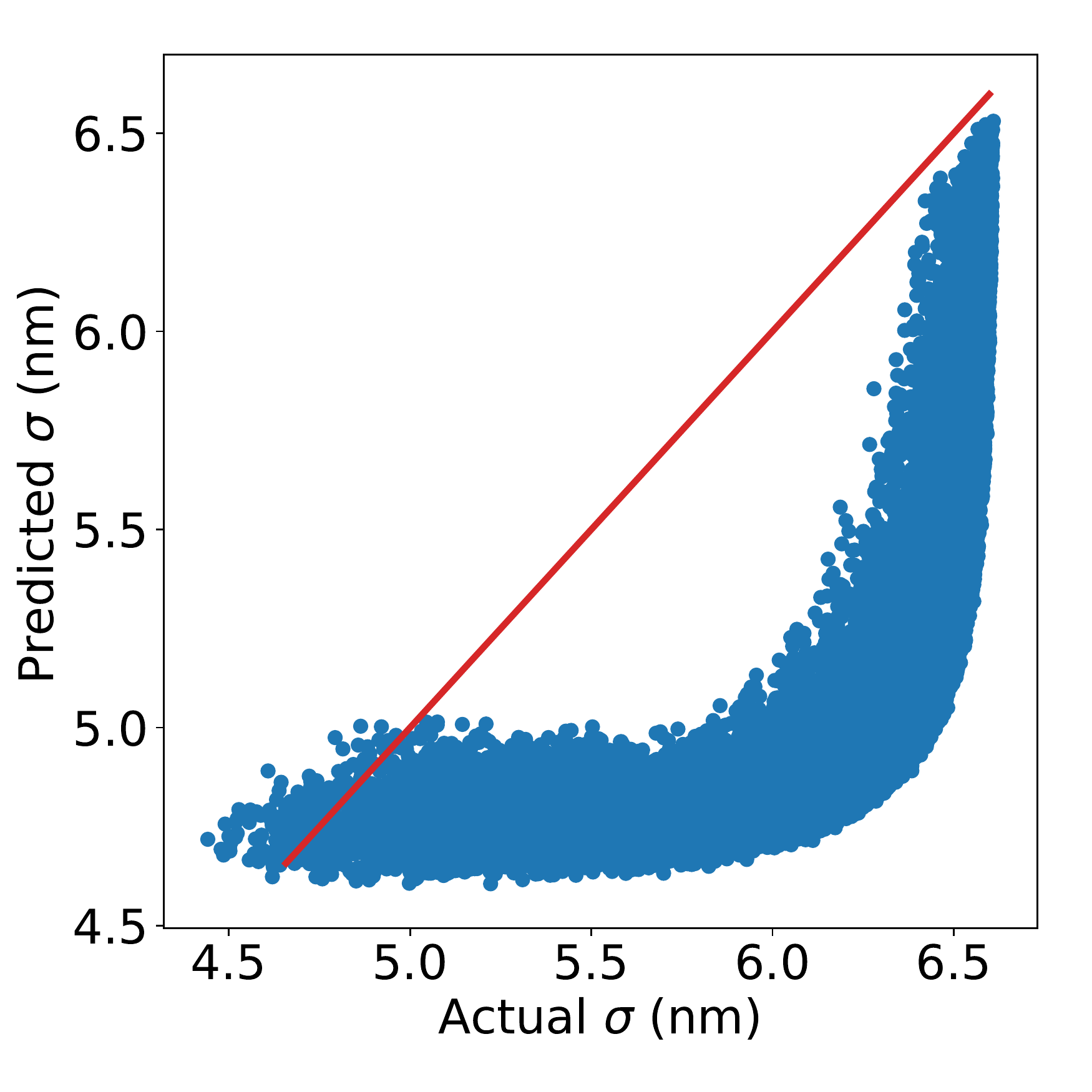}
\label{fig:bigerror}
}
\subfigure[QMC sampling]{
\includegraphics[width=0.45\textwidth]{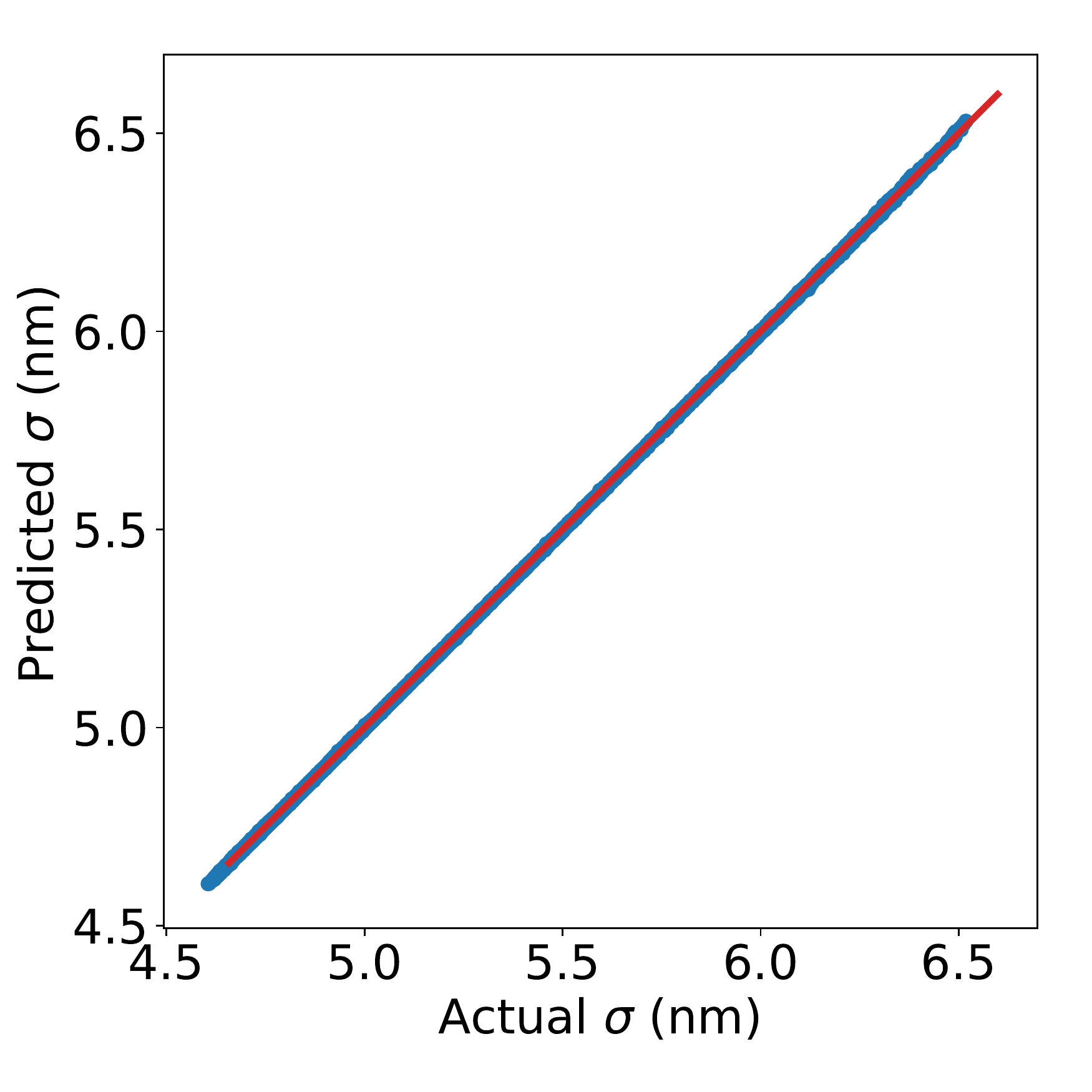}
\label{fig:smallerror}}
\caption{Generalization error of the trained neural network in 2D when the number of random variables is $10$ and $1024$ points are chosen. The relative $L^\infty$ error is $30.561\%$ when uniformly distributed points are used and the relative $L^\infty$ error is $0.237\%$ when  points generated by Sobol sequence are used. {The training data is visualized as a line and the testing data is visualized as scattering points.}}
\end{figure}

In the sampling stage of data preparation, a large $M$ is needed to ensure that the extracted function of EDL has the desired accuracy. There are two classical choices: uniform sampling and random sampling. For uniform sampling, $M$ grows exponentially fast with respect to $K_1$ and $K_2$. For example, in the 3D case, if $K_1=K_2=5$ and points are uniformly distributed for each random variable, the size of training data set is shown in Table \ref{tbl:uniformly}.
For the simulations in our work, at least three orders of magnitude reduction in the size of data set is found for QMC sampling strategy. Fig. \ref{fig:bigerror} and Fig. \ref{fig:smallerror} show the huge advantage of QMC sampling over uniform sampling. For the same size of training data set, the relative $L^\infty$ error is $30.561\%$ and $0.237\%$, implying more than two orders of magnitude improvement in the prediction accuracy.

\begin{figure}[ht]
\centering
\subfigure[Uniform sampling]{
\label{fig:uniformlysampling}
\includegraphics[width=0.45\textwidth]{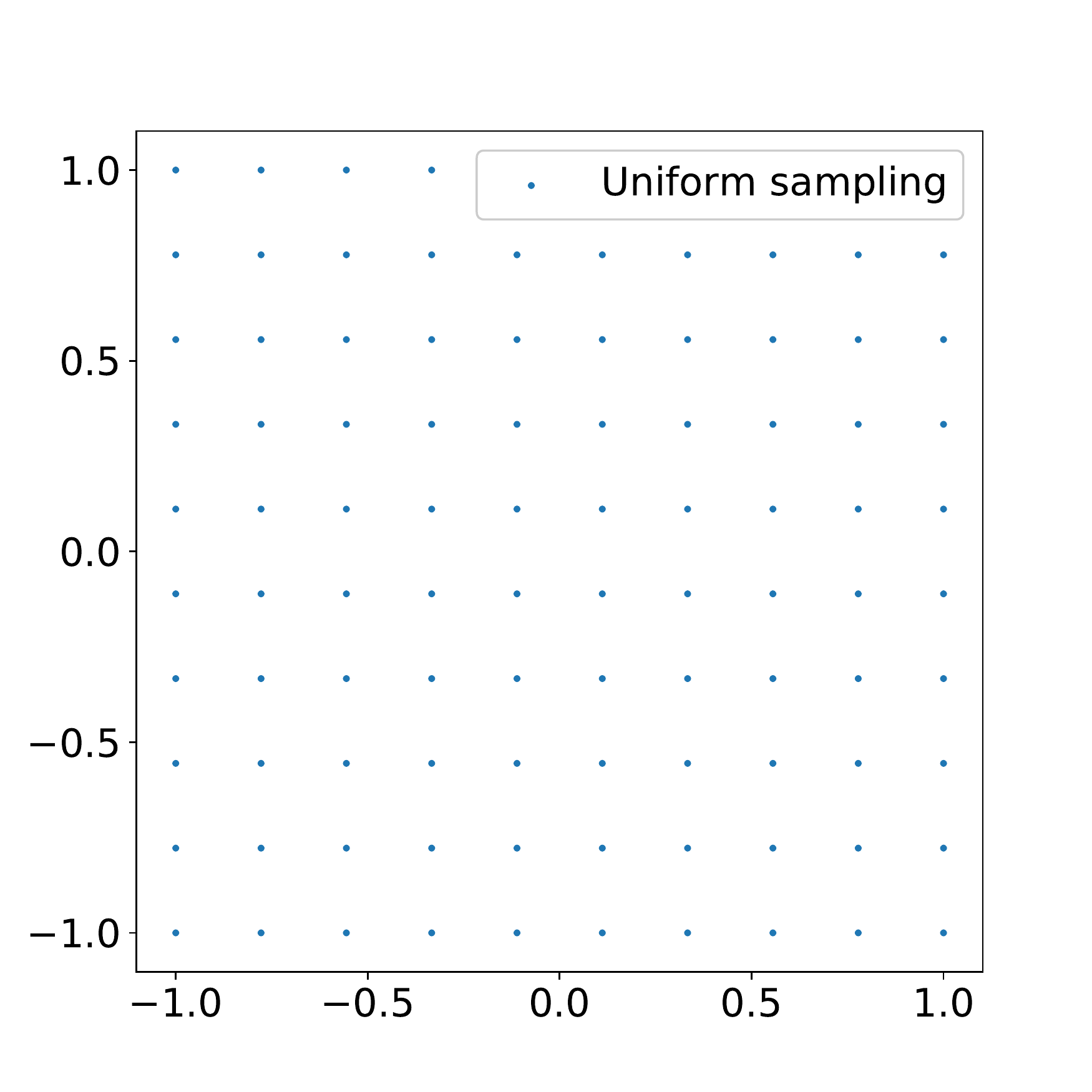}
}
\subfigure[QMC sampling]{
\label{fig:QMCsampling}
\includegraphics[width=0.45\textwidth]{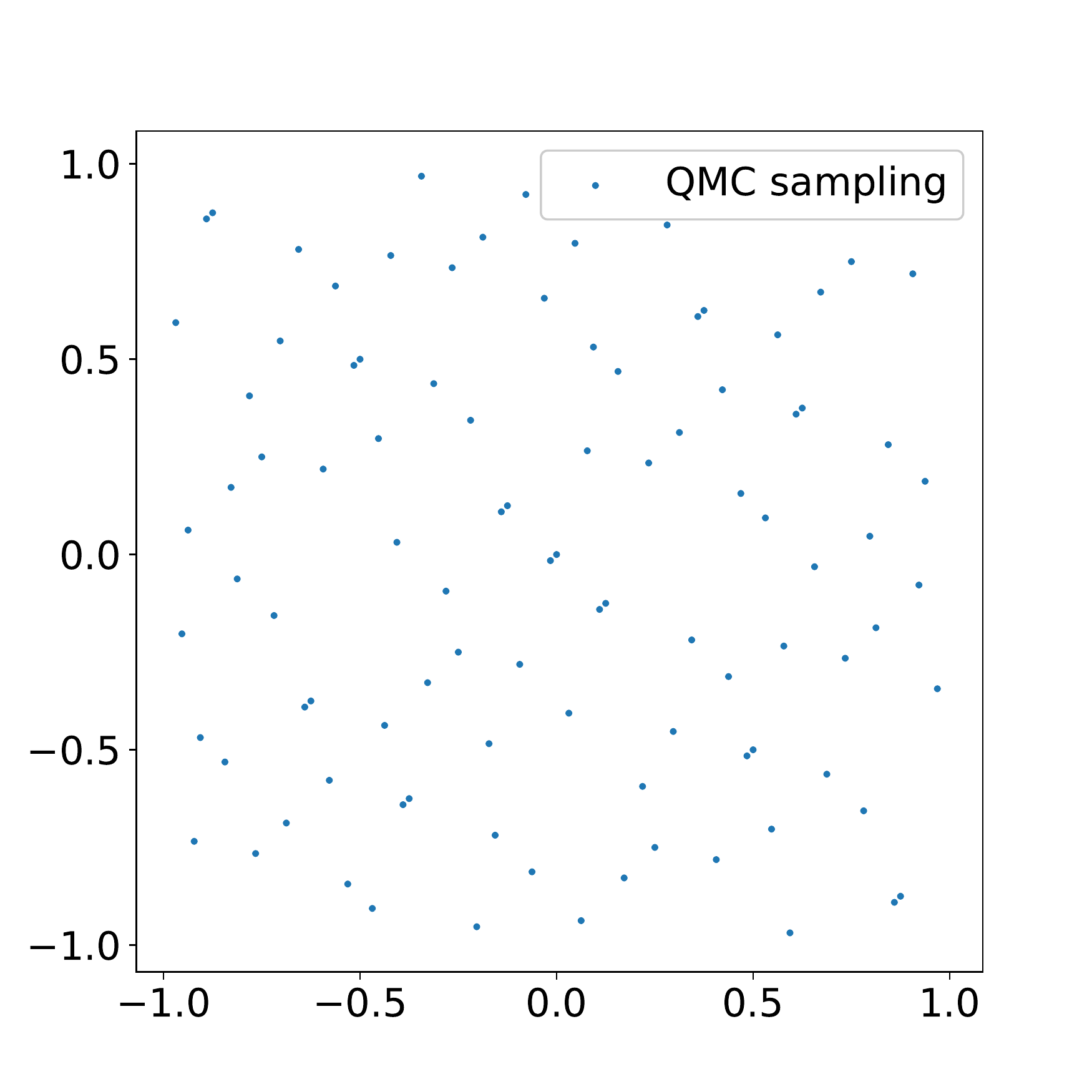}
}
\caption{Comparison of two sampling method. (a) Uniform sampling for two random variables with $100$ points; (b) Quasi-Monte Carlo sampling (Sobol sequence) for two random variables with $100$ points. }\label{fig:sampling method}
\end{figure}

\begin{table}
\centering
	\caption{Size of training data set for uniform sampling.}
	\label{tbl:uniformly}
	\begin{tabular}{|l|l|l|l|l|}
		\hline
		Number of points in each dimension & $2$ & $3$ & $5$ & $9$\\
		\hline
		Size of training data set & $1024$ & $59049$ & $9765625$ & $3486784401$\\
		\hline
	\end{tabular}
\end{table}
Fig. \ref{fig:uniformlysampling} plots the points by uniform sampling when $K_1=K_2=1$ (two random variables). Clearly such a sampling strategy has the curse of dimensionality.
\par
On the other hand, if random sampling is used, then we do not have this issue. However, MC method has poor accuracy $\sim O(\frac{1}{\sqrt{M}})$. At least millions of data are needed for training. Meanwhile, for each datum, an inverse problem with the diffusion equation model over a curved domain in 3D has to be solved. These together make the network training prohibitively expensive. Fortunately, compared to uniform sampling and MC sampling, QMC sampling provides the best compromise between accuracy and efficiency. It overcomes the curse of dimensionality and has high accuracy~\cite{niederreiter1992random, Russel1998, dick2013high}. Specifically, we use Sobol sequence to generate points over the (high-dimensional) random space. Fig. \ref{fig:QMCsampling} plots the points generated by Sobol sequence, which is a deterministic way to generate points with better approximation accuracy. QMC sampling has accuracy $\sim O(\frac{(\log M)^{\mathrm{dim}}}{M})$~\cite{niederreiter1992random, Russel1998, dick2013high}, which reduces the size of training data set by orders of magnitudes in comparison with MC method. The size of data in QMC method grows merely linearly fast with respect to the number of random variables.

\subsection{ResNet}
\label{sec:detailofDNN}
ResNet~\cite{He2015} is used to approximate $\sigma_{\theta(\omega_1),\theta(\omega_2)}$. A ResNet consists of a series of blocks. One block is given in Fig. \ref{fig:schematic} with two linear transformations, two activation functions, and one short cut.The $i$-th block can be expressed as
\begin{equation}\label{equ:ith block}
  t= f_i(s)= g(W_{i,2}\cdot g(W_{i,1}\cdot s +b_{i,1})+b_{i,2})+s.
\end{equation}
Here $s,t\in R^{m}$ are input and output of the $i$-th block, and weights $W_{i,j}\in R^{m\times m}, b_{i,1}, b_{i,2}\in R^{m}$. Sigmoid function, which can be expressed in
\begin{equation*}%\label{equ:activation function}
g(x)=\frac{1}{1+\exp(-x)},
\end{equation*}
is chosen as the activation function to balance training complexity and accuracy.

The last term in \eqref{equ:ith block} is called the shortcut connection or the residual connection. Advantages of using it are
\begin{enumerate}[1)]
\item It can solve the notorious problem of vanishing/exploding gradients automatically. {The vanishing gradient problem means that in some cases the gradient will be vanishingly small, effectively preventing the weight from changing its value. In the worst case, this may completely stop the neural network from further training. Also, when activation functions are used whose derivatives can take on larger values, one risks encountering the related exploding gradient problem.}
\item Without adding any parameters or computational complexity, the shortcut connection performing as an $Identity$ mapping can resolve the degradation issue (with the network depth increasing, accuracy gets saturated and then degrades rapidly).
\end{enumerate}
The fully $n$-layer network can be expressed as
\begin{equation*}\label{equ:fully network}
  f_w(x)= f_n \circ f_{n-1} \cdots \circ f_1(x),
\end{equation*}
where $w$ denotes the set of parameters in the whole network. Note that the input $x$ in the first layer is in $R^{\mathrm{dim}}$ and the output of the whole structure $\sigma(\theta(\omega_1),\theta(\omega_2))$ is in $R^1$. To deal with the problem, we apply two linear transformations on both $x$ before putting it into the ResNet structure and on the output of the ResNet structure. For example, we choose $m=30,n=6$ in the 3D model. Both $\theta(\omega_1)$ and $\theta(\omega_2)$ have $5$ random variables, and thus $\mathrm{dim}=10$. Therefore, we apply two linear transforms: one from a $10$ dimensional vector to a $30$ dimensional vector and the other from a $ 30$ dimensional vector to $1$ dimensional vector before and after the ResNet structure. Parameters in these linear transforms also need to be trained.

The loss function we use is the MSE between the actual EDL $\sigma_{\theta(\omega_1),\theta(\omega_2)}$ given by the diffusion equation model and the predicted EDL $\tilde\sigma(\theta(\omega_1)[j],\theta(\omega_2)[j])$ given by the ResNet
\begin{equation}\label{eqn:MSE}
  MSE= \frac{1}{M}\sum_{j=1}^{M} \left(\sigma_{\theta(\omega_1)[j],\theta(\omega_2)[j]}-\tilde\sigma(\theta(\omega_1)[j],\theta(\omega_2)[j])\right)^2,
\end{equation}
where $\theta$ represents the parameter set in the ResNet, $j$ is the $j$-th sample, and M is the size of training data set.
\par
Define the relative $L^{\infty}$ error of EDL as
\begin{equation}
Error =  \max_{1\leq j\leq M} \dfrac{\left|\sigma_{\theta(\omega_1)[j],\theta(\omega_2)[j]}-\tilde\sigma(\theta(\omega_1)[j],\theta(\omega_2)[j])\right|} {\sigma_{\theta(\omega_1)[j],\theta(\omega_2)[j]}},
\end{equation}
which will be used to quantify the approximation accuracy of DL.

%-----------------------------------------------------------------------------

\section{Results and discussion}
\label{Sec:Result and Discussion}
\subsection{Accuracy check and training data set}
\textbf{\textit{2D}}
First, we focus on the 2D problem with only one realization, i.e., only one $d=10$ and $N=1$. PL data are generated when $\sigma=10$ without any randomness. Accuracy of the trained neural network in terms of size of the training set is recorded in Table \ref{tbl:error in 2D}.
\begin{table}
\centering
\caption{Generalization error of the trained neural network model for a random field with different decay rates in 2D.}\label{tbl:error in 2D}
\begin{tabular}{|l|l|l|l|l|}
\hline
Size of training data set & $9$ & $25$ & $81$ \\
\hline
Error ($\beta=-2$) & $1.638\%$ & $0.04466\%$ & $0.00209\%$\\
\hline
Error ($\beta=-1$) & $0.101\%$ & $0.00795\%$ & $0.00234\%$\\
\hline
Error ($\beta=0$) & $0.765\%$ & $0.105\%$ & $0.0180\%$\\
\hline
\end{tabular}
\end{table}
From the results, we can find that a random field with the slower decay rate ($\beta=0$) is more difficult to be trained when uniform sampling is used.

\par
\begin{figure}[ht]
	\centering \includegraphics[width=0.55\textwidth]{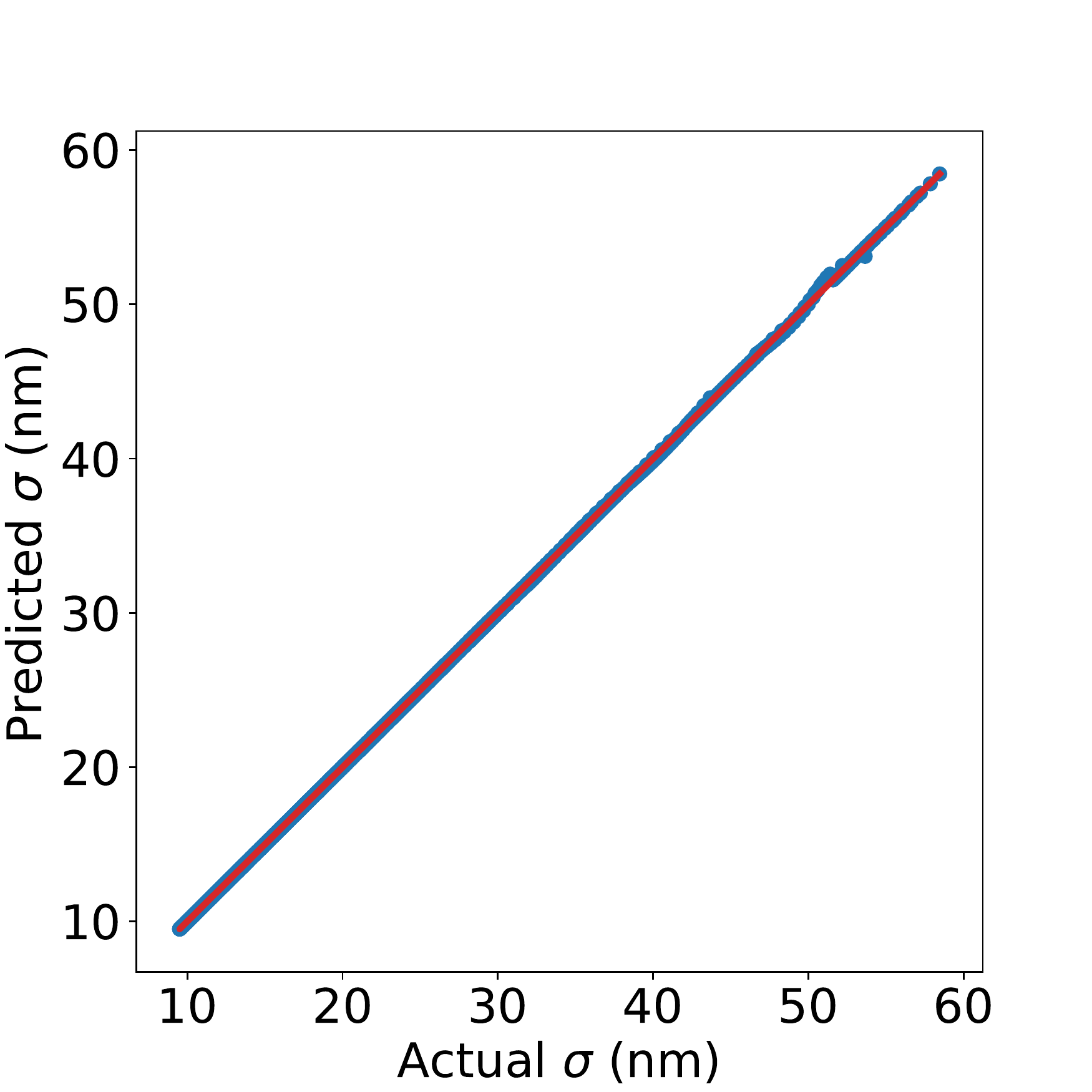}
	\caption{Generalization error of the trained neural network for random variables ranging over $[-5,5]$ with $1$ photoluminescence datum in 2D. The relative $L^\infty$ error is $1.071\%$. The training data is visualized as a line and the testing data is visualized as scattering points.}\label{fig:bigrandomnesserror}
\end{figure}
\begin{figure}[ht]
\centering
\subfigure[$\beta=-2$]{
\includegraphics[width=0.3\textwidth]{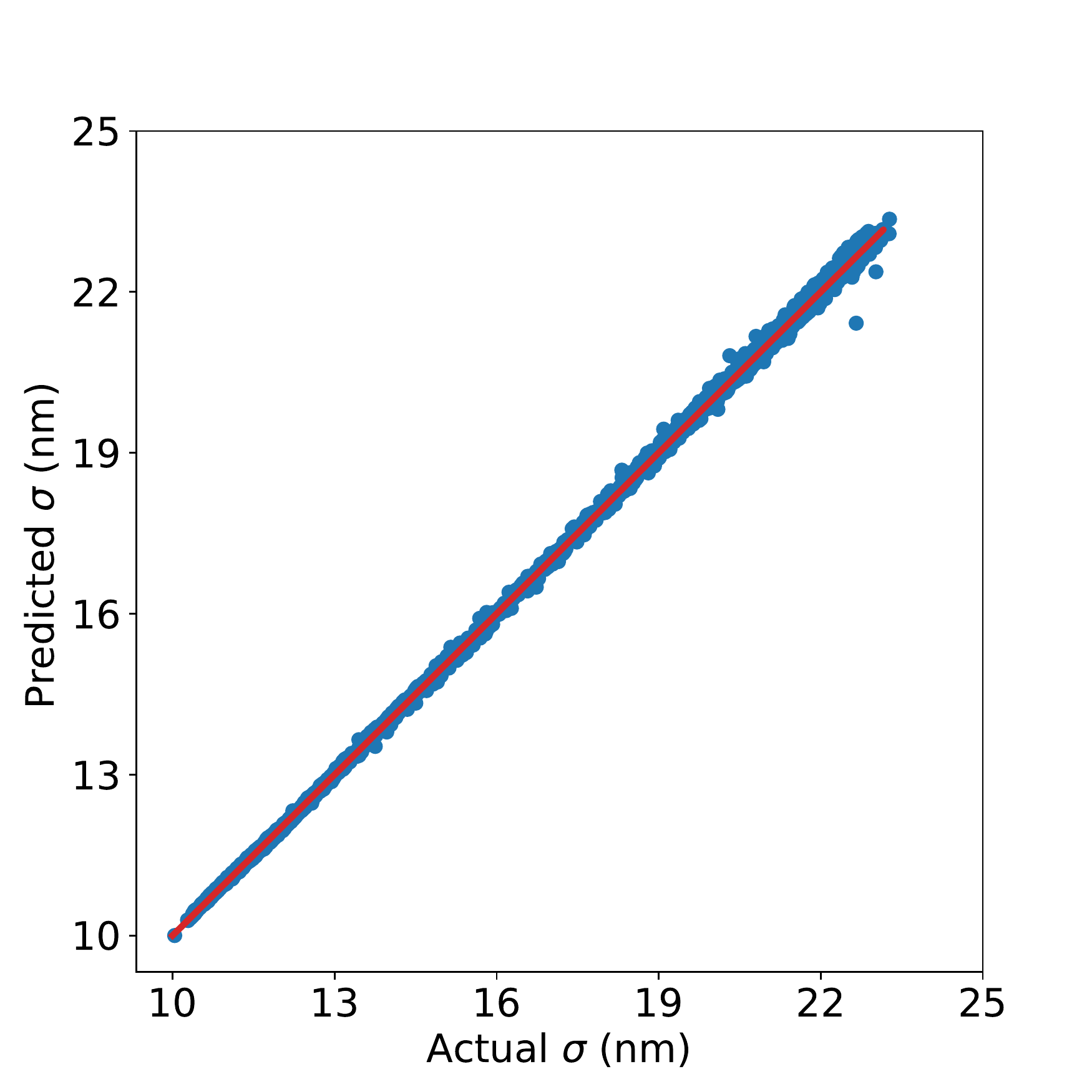}
\label{fig:beta2realisticerror}
}
\subfigure[$\beta=-1$]{
\includegraphics[width=0.3\textwidth]{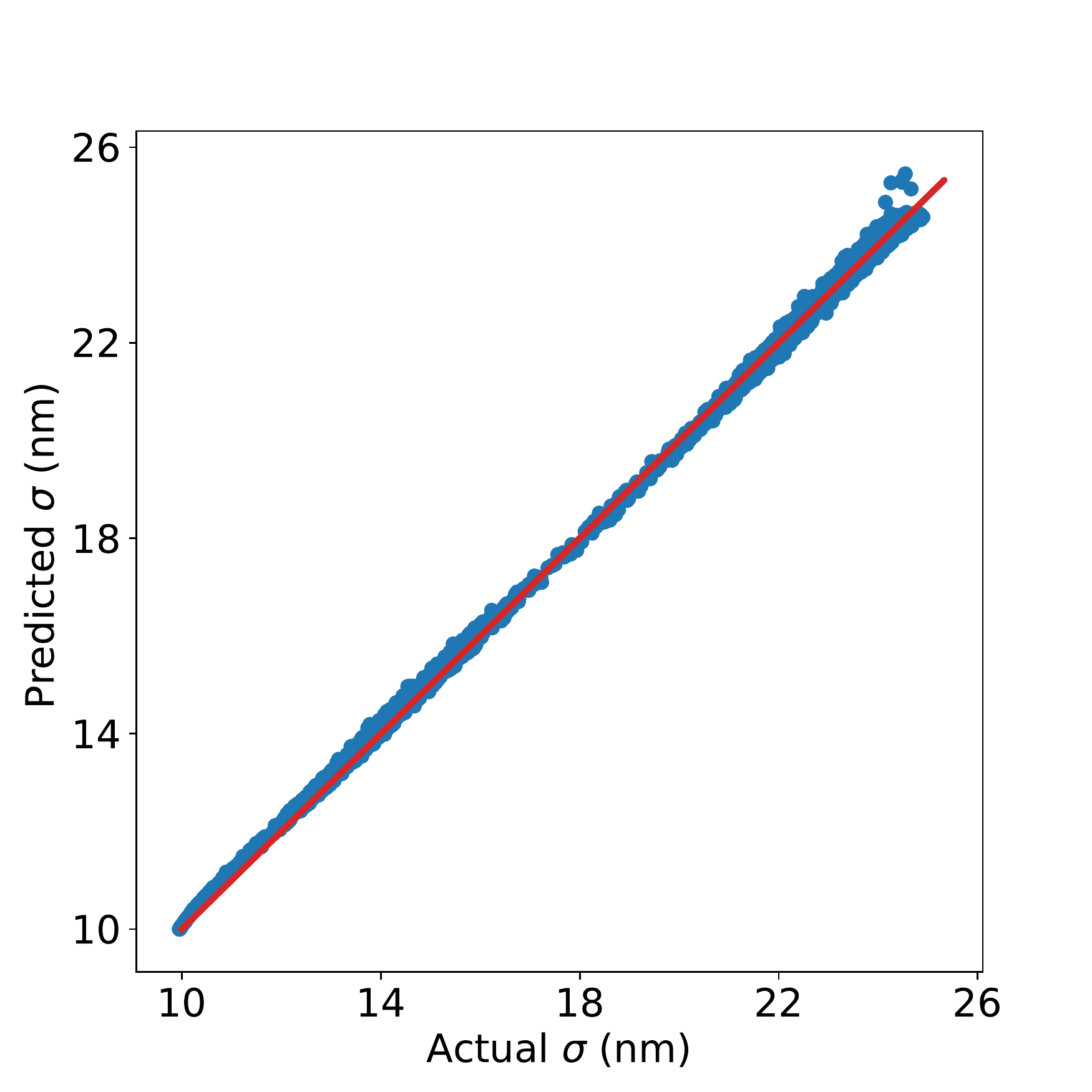}
\label{fig:beta1realisticerror}
}
\subfigure[$\beta=0$]
{
\includegraphics[width=0.3\textwidth]{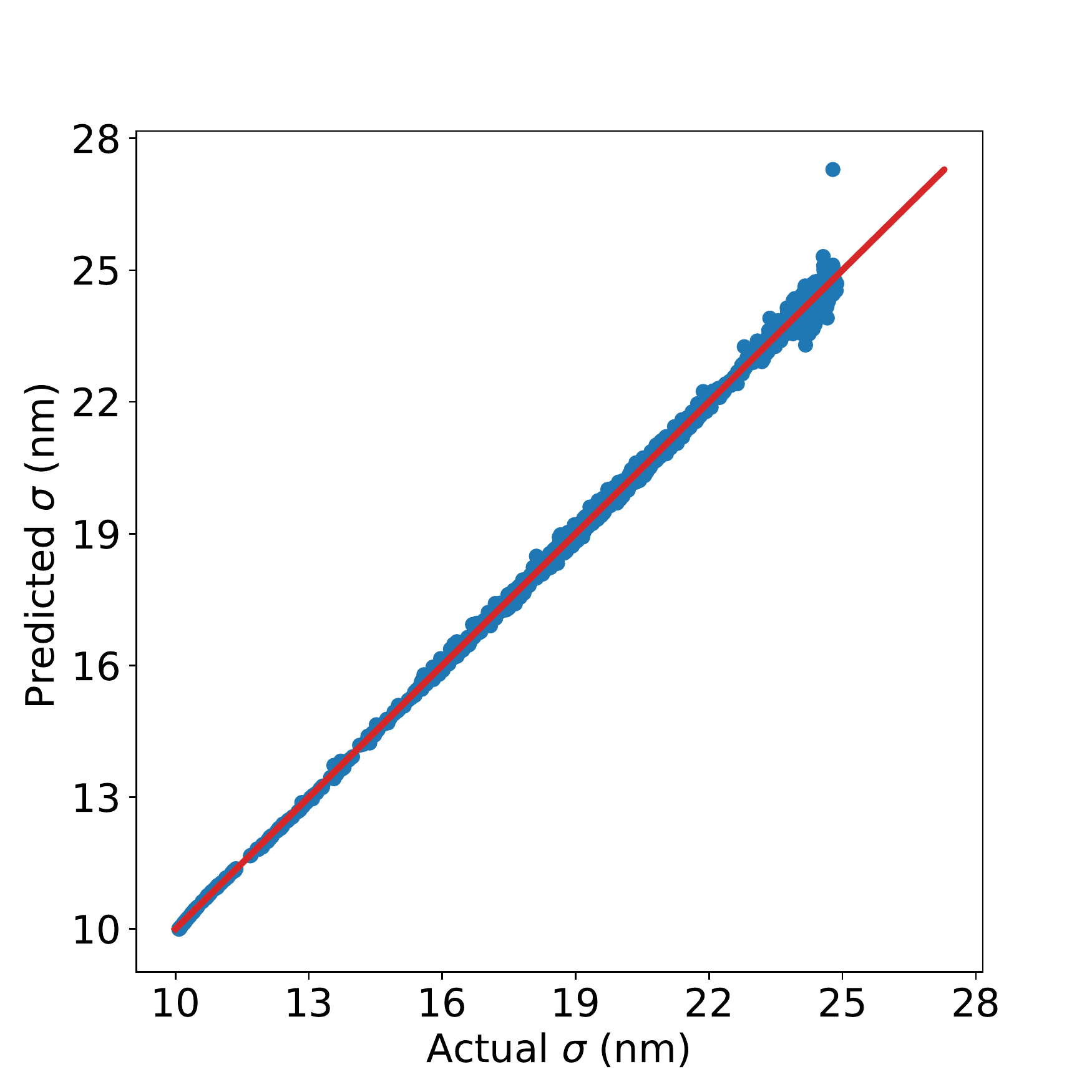}
\label{fig:beta0realisticerror}
}
\caption{Generalization error of the trained neural network for random variables ranging over {$[-5,5]$} with $6$ photoluminescence data and $\beta=0$ in 2D. (a) The relative $L^\infty$ error is $5.784\%$ when $\beta=-2$; (b) The relative $L^\infty$ error is $3.327\%$ when $\beta=-1$; (c) The relative $L^\infty$ error is $9.184\%$ when $\beta=0$. {The training data is visualized as a line and the testing data is visualized as scattering points.}}\label{fig:betarealisticerror}
\end{figure}

In the literature, a asymptotics-based method has been proposed~\cite{Chen2019894}, which only works well for random interfaces with small magnitudes. The proposed method works for random interfaces with large magnitudes. For example, consider $\theta(\omega)$ with $2$ random variables ranging over $[-5,5]$ and $\beta = 2$, i.e., the dimensionless magnitude of perturbation is about $0.5$, therefore the asymptotics-based method has poor accuracy in this case unless enough terms are used in the asymptotic expansion. {For example, we use the asymptotics-based method with up to the second order terms \cite{Chen2019894, Chen2020} and the average error is over $25\%$.} However, for the QMC-machine learning method, the relative $L^{ \infty }$ error is $1.071\%$; see Fig. \ref{fig:bigrandomnesserror}.
For a random field with $10$ random variables and $5$ realizations $d=[10, 15 , 20, 30, 40,50]$, generalization errors of the trained neural network are plotted in Fig. \ref{fig:betarealisticerror} for $\beta=-2, -1, 0$, respectively. Again, the dimensionless magnitude of perturbation is about $0.5$, our method still works with error smaller than $10\%$, which is an acceptable tolerance in experiments.

\begin{figure}[ht]
\centering
\subfigure[$\beta=-2$]{
\includegraphics[width=0.3\textwidth]{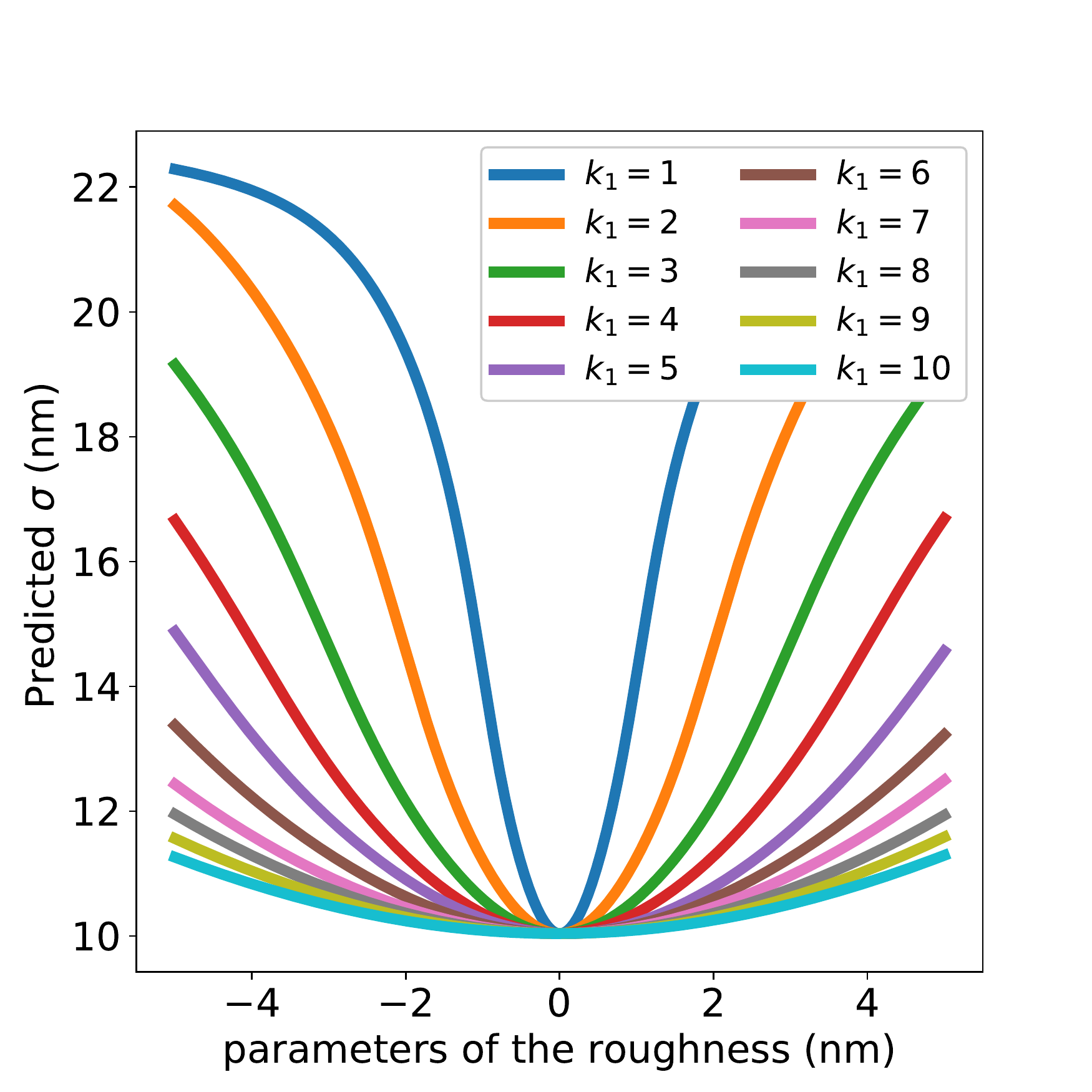}
\label{fig:beta2realistic}
}
\subfigure[$\beta=-1$]{
\includegraphics[width=0.3\textwidth]{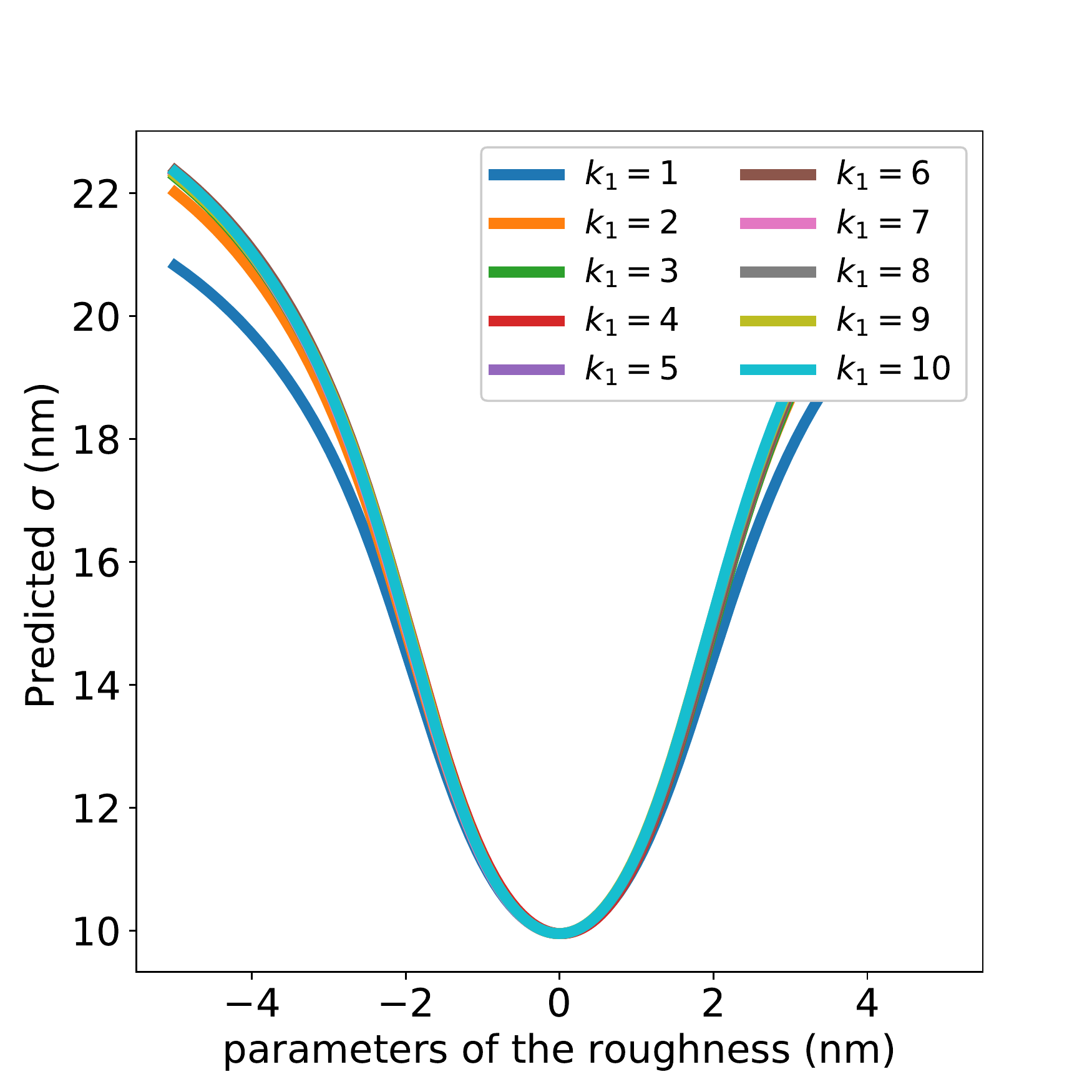}
\label{fig:beta1realistic}
}
\subfigure[$\beta=0$]{
\includegraphics[width=0.3\textwidth]{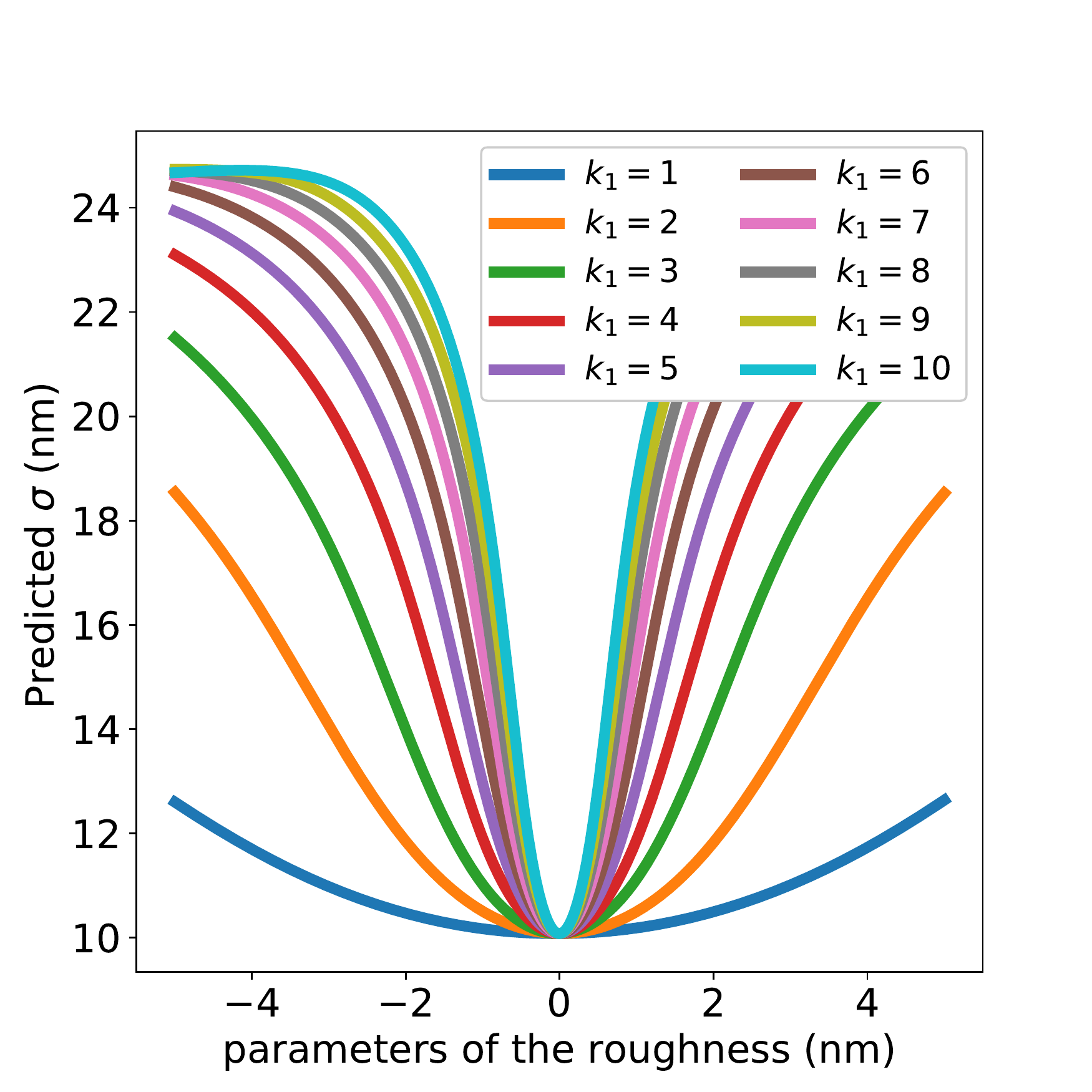}
\label{fig:beta0realistic}
}
\caption{Dependence of exciton diffusion length on random variables in 2D. (a) $\beta=-2$; (b) $\beta=-1$; (c) $\beta=0$.}\label{fig:betarealistic}
\end{figure}

In 2D, a detailed dependence of EDL $\sigma$ on random variables is given in Fig. \ref{fig:betarealistic} for $\beta=-2, -1, 0$, respectively. {In these plots, we visualize EDL on the chosen random variable with all the other parameters fixed to be $0$.} An interesting observation is that the larger the randomness is, the larger the EDL is. This is in contrast with experimental experiences and 3D results in Fig. \ref{fig:lambda0d1}. We attribute this observation to the particularity of dimension, which may deserve further experimental confirmations.

\par
\begin{figure}[ht]
\centering
\subfigure[$\beta=-2$]{
\includegraphics[width=0.28\textwidth]{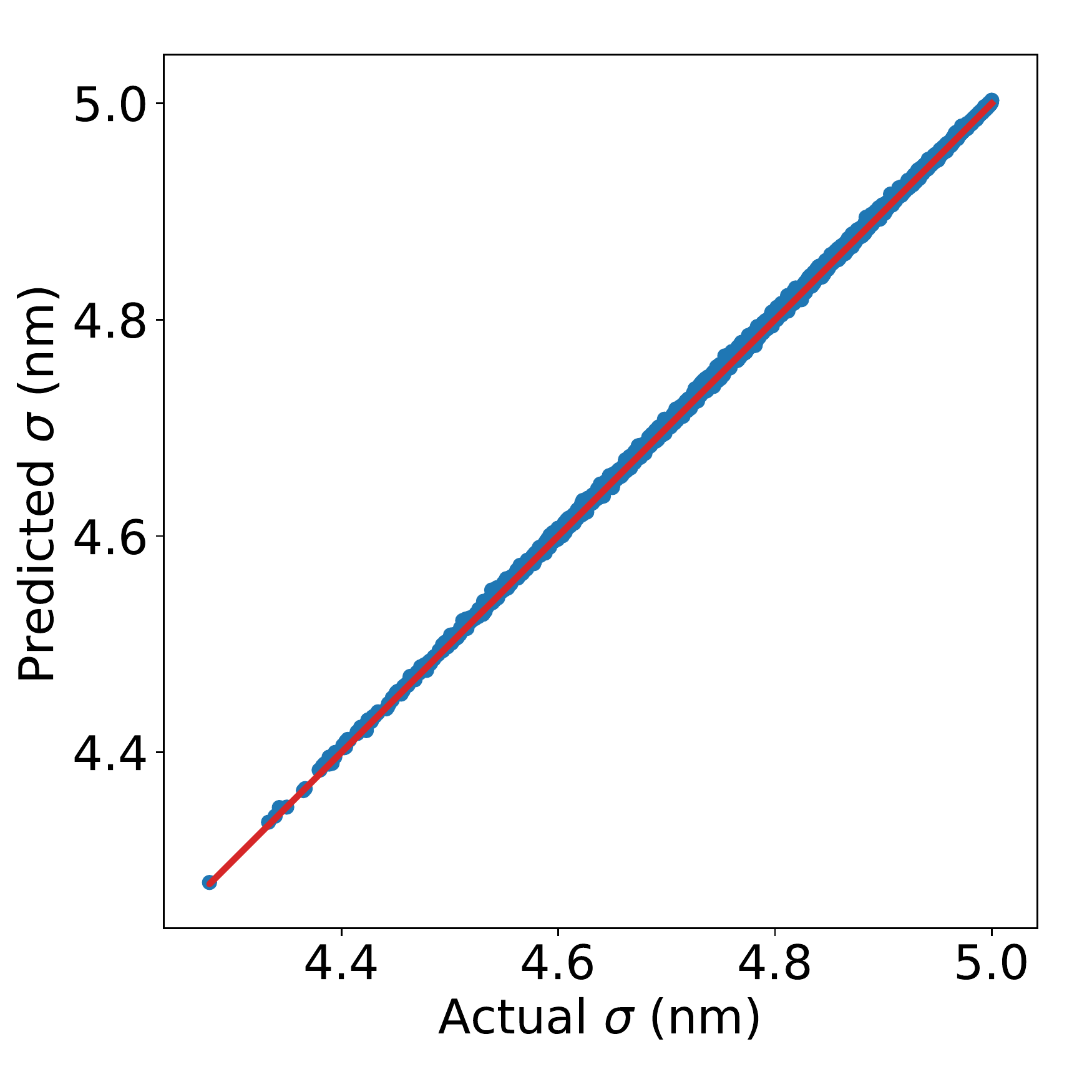}
}
\quad
\subfigure[$\beta=-1$ ]{
\includegraphics[width=0.28\textwidth]{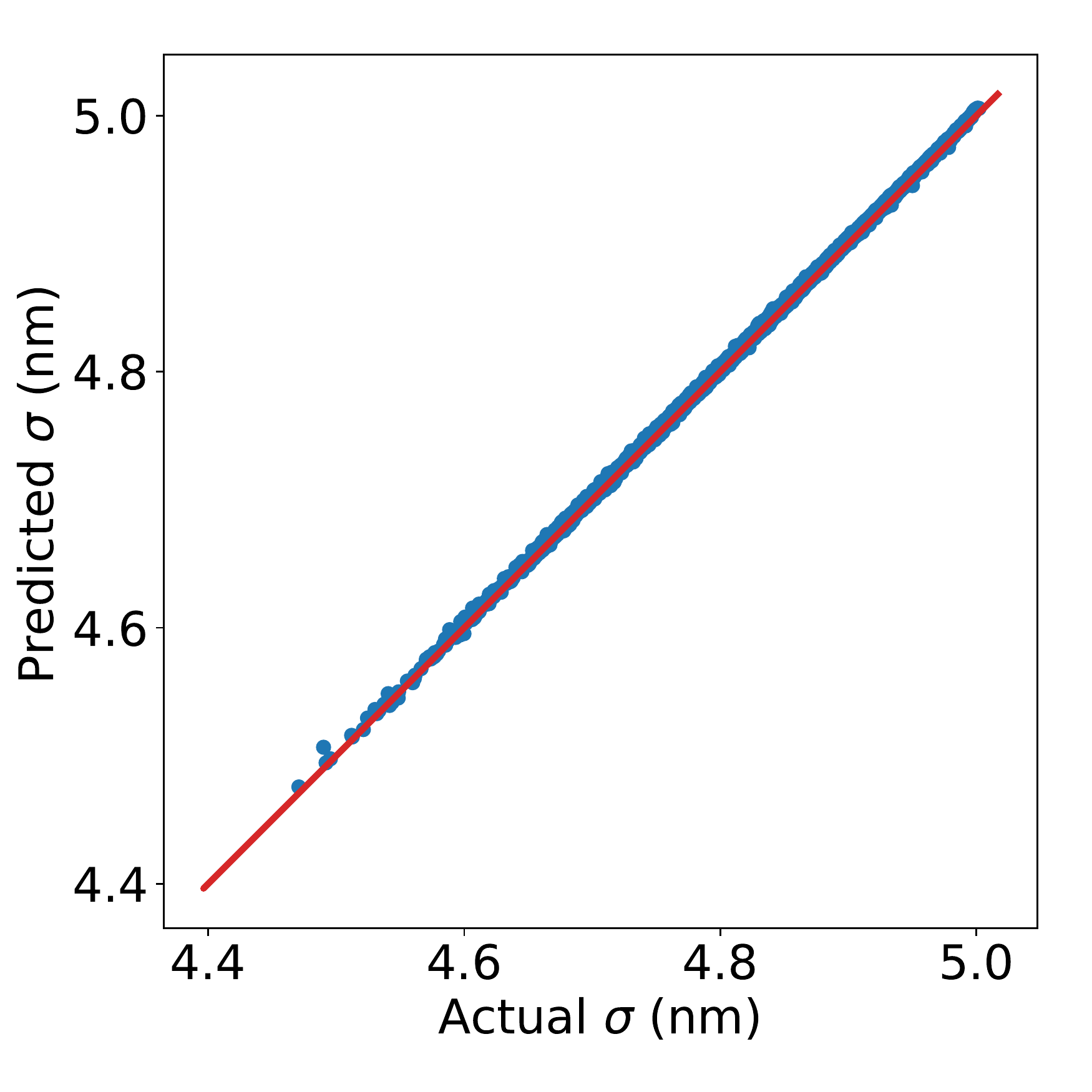}
}
\quad
\subfigure[$\beta=0$]{
\includegraphics[width=0.28\textwidth]{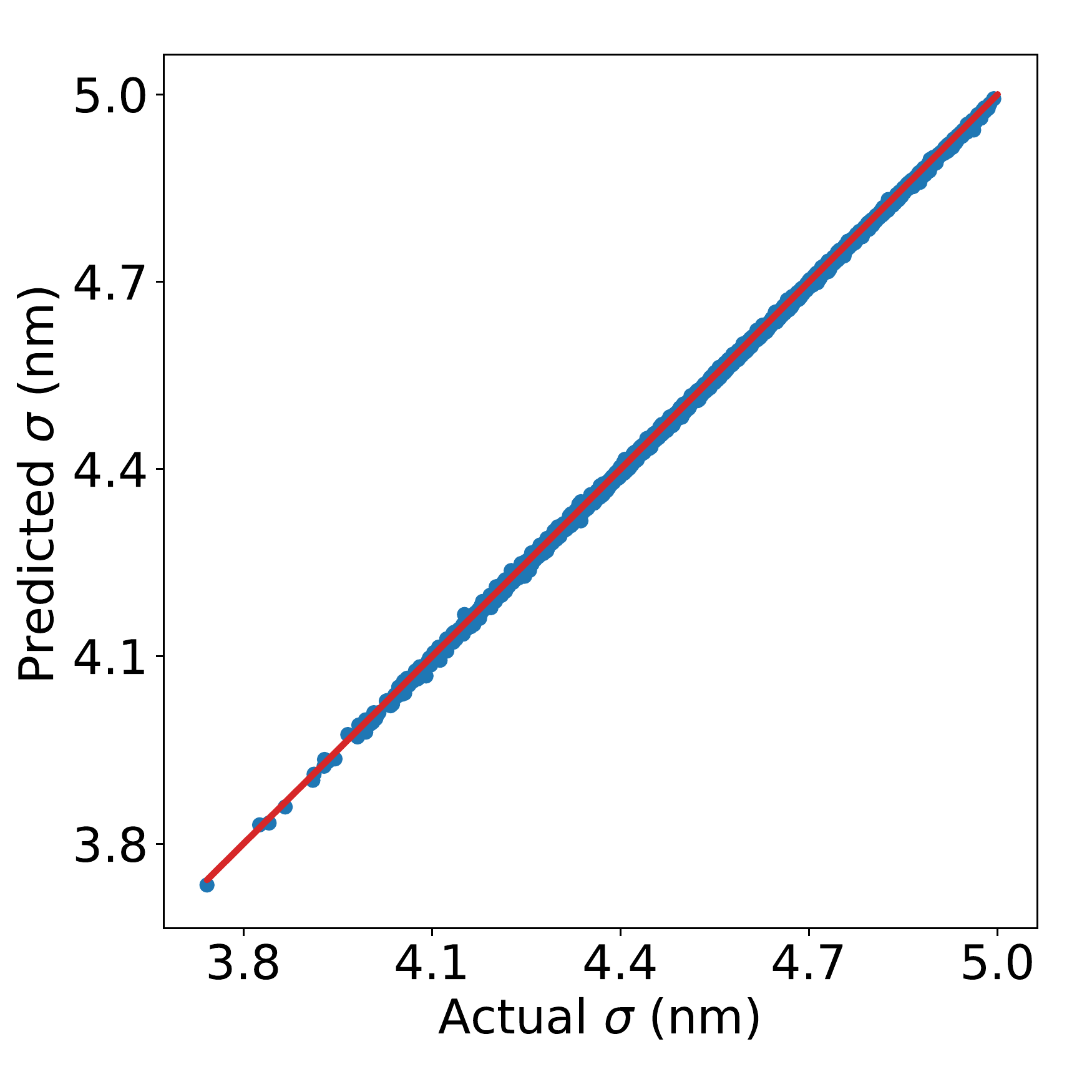}
}
\caption{Uniform generalization error of the trained neural network for exciton diffusion length when $\beta=-2, -1, 0$. Relative $L^\infty$ errors of exciton diffusion length are $0.270\%,0.368\%$ and $0.532\%$, respectively. {The training data is visualized as a line and the testing data is visualized as scattering points.}}\label{fig:actualpredict}
\end{figure}
\begin{figure}[ht]
\centering
\subfigure[$\beta=-2$]{
\includegraphics[width=0.29\textwidth]{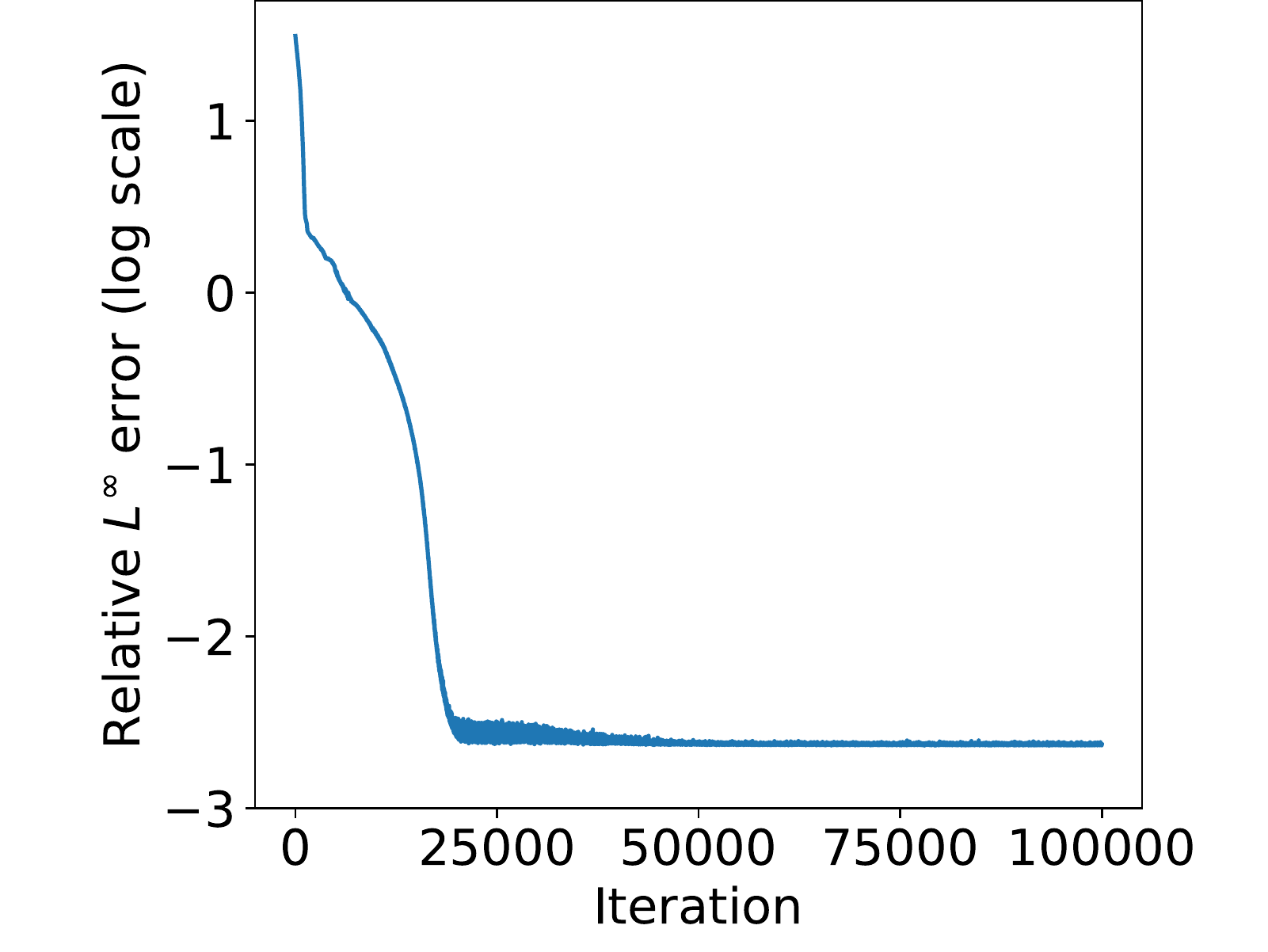}
}
\quad
\subfigure[$\beta=-1$ ]{
\includegraphics[width=0.29\textwidth]{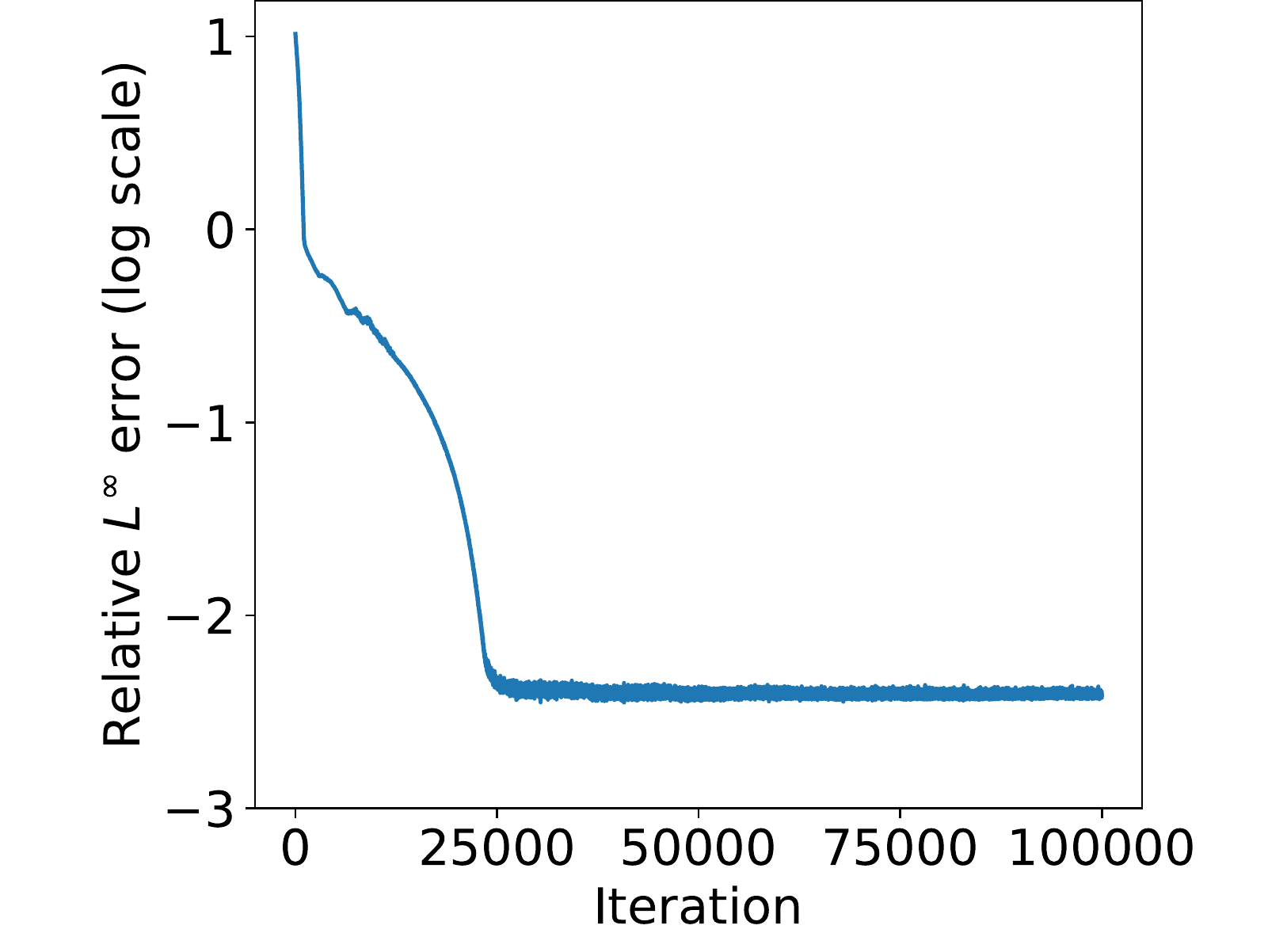}
}
\quad
\subfigure[$\beta=0$]{
\includegraphics[width=0.29\textwidth]{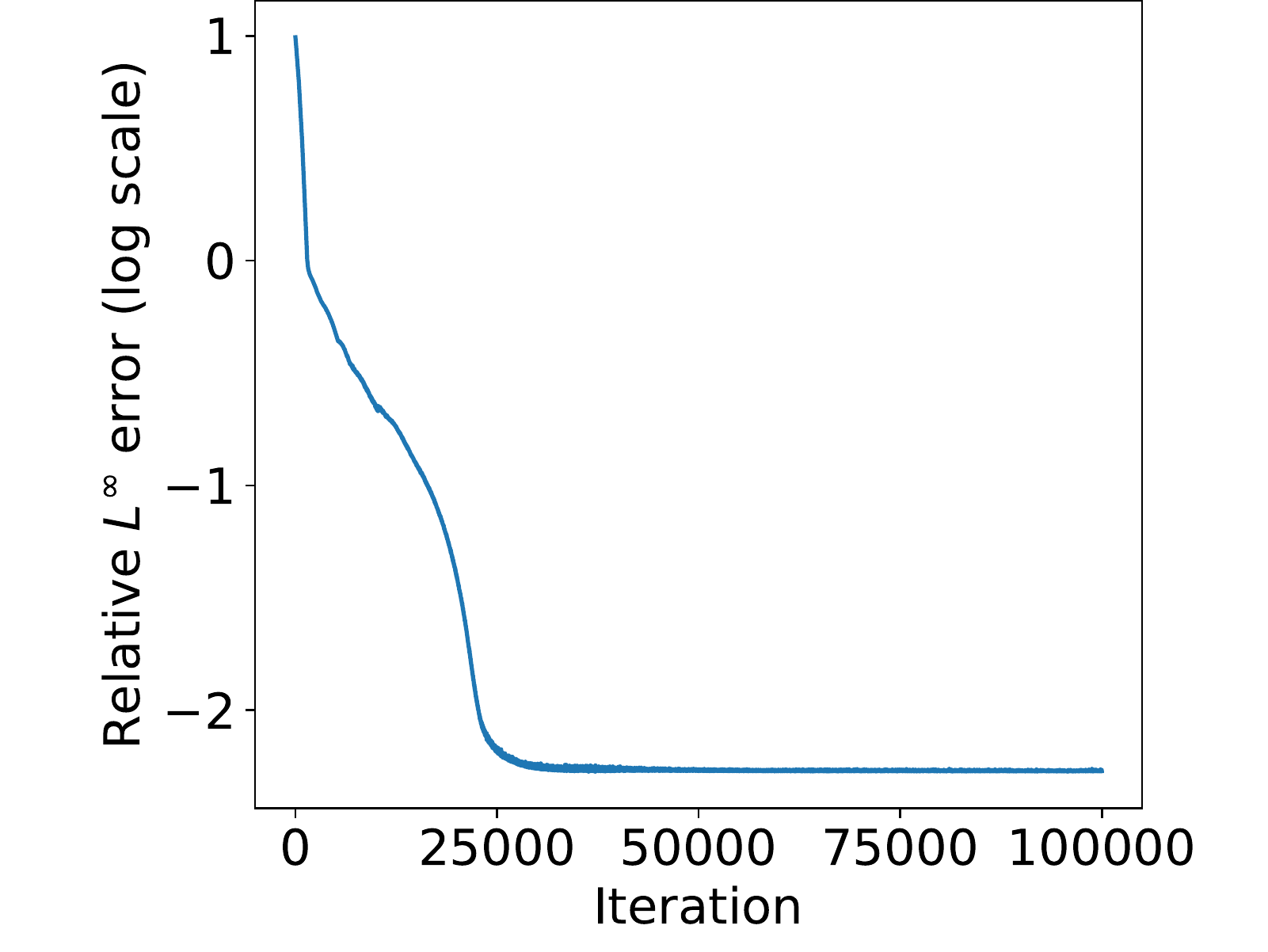}
}
\caption{Relative $L^\infty $ error of exciton diffusion length in terms of iteration number for $\beta = -2, -1, 0$, respectively. }\label{fig:erroriteration}
\end{figure}

\textbf{\textit{3D}}
For the accuracy check in the 3D case, the reference PL data are generated using $5$ realizations with out-of-plane thicknesses $d_i=10,15, 20,25\;$nm and $\sigma=5\;$nm in the absence of randomness. Afterwards, randomness is added with $K_1=K_2=5$, i.e., $\theta(\omega_1)$ and $\theta(\omega_2)$ are arrays with $5$ variables. QMC sampling is used to generate 20000 points with the corresponding EDL obtained by
solving \eqref{eqn:3DInterface} - \eqref{eqn:Inverseproblem}. The first $15000$ data are used as the training set, while the remaining data are used to check
the predictability of the trained neural network; see Fig. \ref{fig:actualpredict}.
Relative $L^\infty$ errors of EDL are $0.270\%,0.368\%$ and $0.532\%$ for $\beta=-2, -1, 0$, respectively. {For completeness, we also plot the convergence history of the training process in terms of the iteration number in Fig. \ref{fig:erroriteration}.} It is known that the random field is closer to the white noise when $\beta=0$ and thus is more difficult to be trained. However, uniform generalization errors for three different scenarios are observed, implying the robustness of trained neural network. Moreover, the size of training data set
is small in the sense that only linear growth with respect to the dimension of random variables is observed, in contrast to other sampling techniques which either have the curse of dimensionality or low accuracy.

\subsection{Information extraction}
The trained neural network fits a high-dimensional function for EDL in terms of surface roughness. Rich information can be extracted based on the fitted function. We demonstrate this using three examples.
\par
\textbf{\textit{Modeling error}}
Expectations of EDL in 3D are recorded in Table \ref{tbl:result} for $\beta = -2, -1, 0$. PL data are generated using the 1D model with the reference EDL $5\;$nm. When $\beta=-2$, the EDL is close to $5\;$nm, which implies the equivalence between the 3D model and the 1D model. However, when $\beta=0$, the EDL is clearly away from $5\;$nm. We attribute this difference to the modeling error between the 1D model and the 3D model with a surface roughness characterized by \eqref{eqn:3DInterface} with $\beta=0$. So far, the 1D model is largely used in the literature to extract the EDL~\cite{Pettersson1999487,Lin2014280,Guideetal:2013,Chen2016754}. The main assumption underlying the modeling is the high crystalline order of the organic material. When $\beta=-2$, long-range ordering exists in the random interface, which implicitly connects with the crystalline ordering of the material. Therefore, in this case, the 3D model and the 1D model are equivalent. However, when $\beta=0$, only short-range ordering exists. As a consequence, the 3D model and the 1D model are not equivalent any more. {This has been verified to be true over a range of EDLs.} Given a surface roughness from the experimental measurement, we can fit a function of form \eqref{eqn:3DInterface} using discrete Fourier transform, from which we can get the decay rate $\beta$ and thus decide whether the 1D model is adequate or not. It is worth mentioning that similar results are observed in 2D using the asymptotics-based approach~\cite{Chen2019894}.
\begin{table}
\centering
	\caption{Expectations of exciton diffusion length in 3D for different surface roughness. The reference value is $5\;$nm. }
	\label{tbl:result}
	\begin{tabular}{lll}
		\hline
		$\beta=-2$  & $\beta=-1$ & $\beta=0$   \\
		\hline
		$4.986\;$nm & $4.842\;$nm & $4.566\;$nm  \\
		\hline
	\end{tabular}
\end{table}
\par
\begin{figure}[ht]
	\centering
	\subfigure[$\beta=-2$]{
		\includegraphics[width=0.48\textwidth]{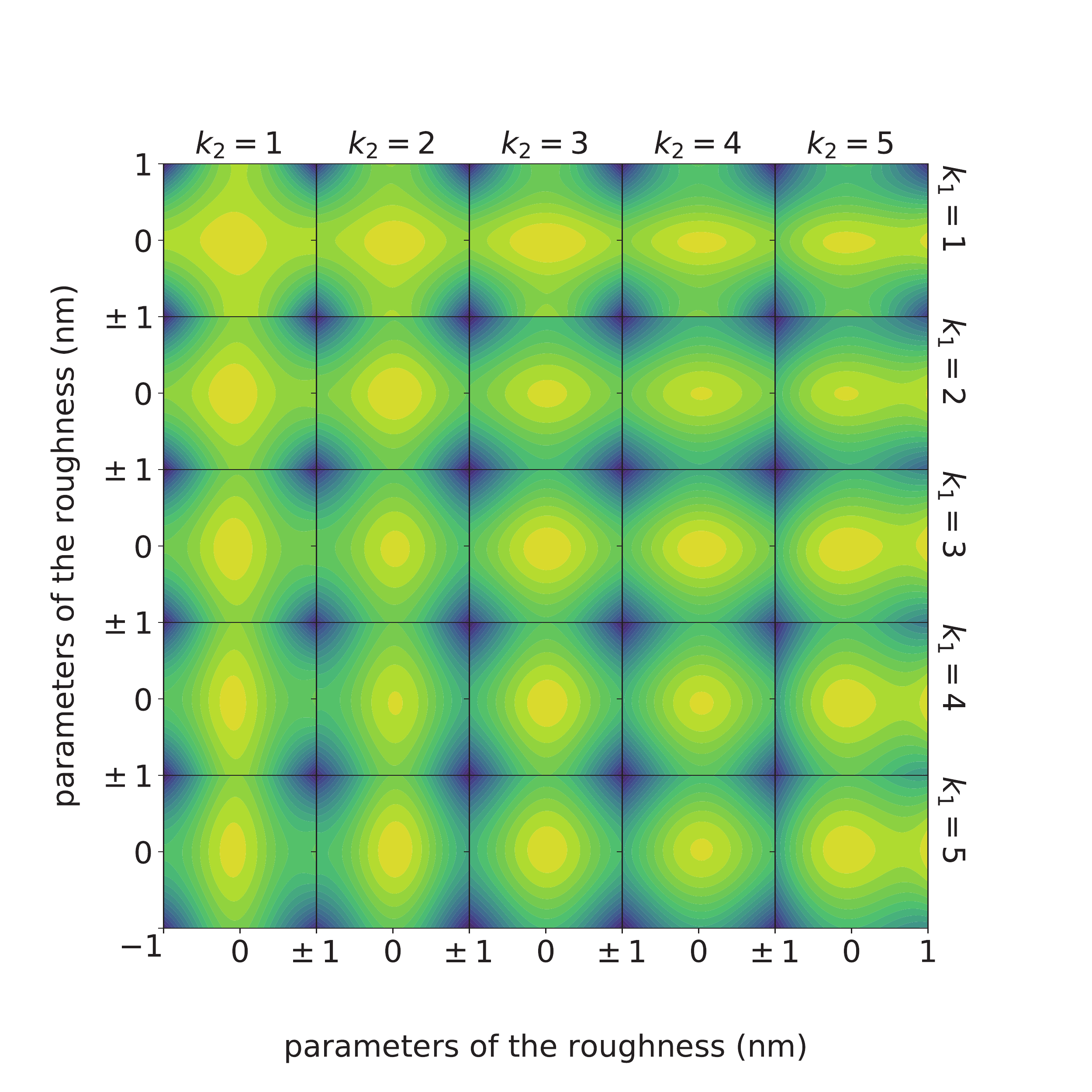}
	}
%	\subfigure[$\beta=-1$ ]{
%		\includegraphics[width=0.4\textwidth]{Figure//lambda1contour-eps-converted-to.pdf}
%	}
	\subfigure[$\beta=0$]{
		\includegraphics[width=0.48\textwidth]{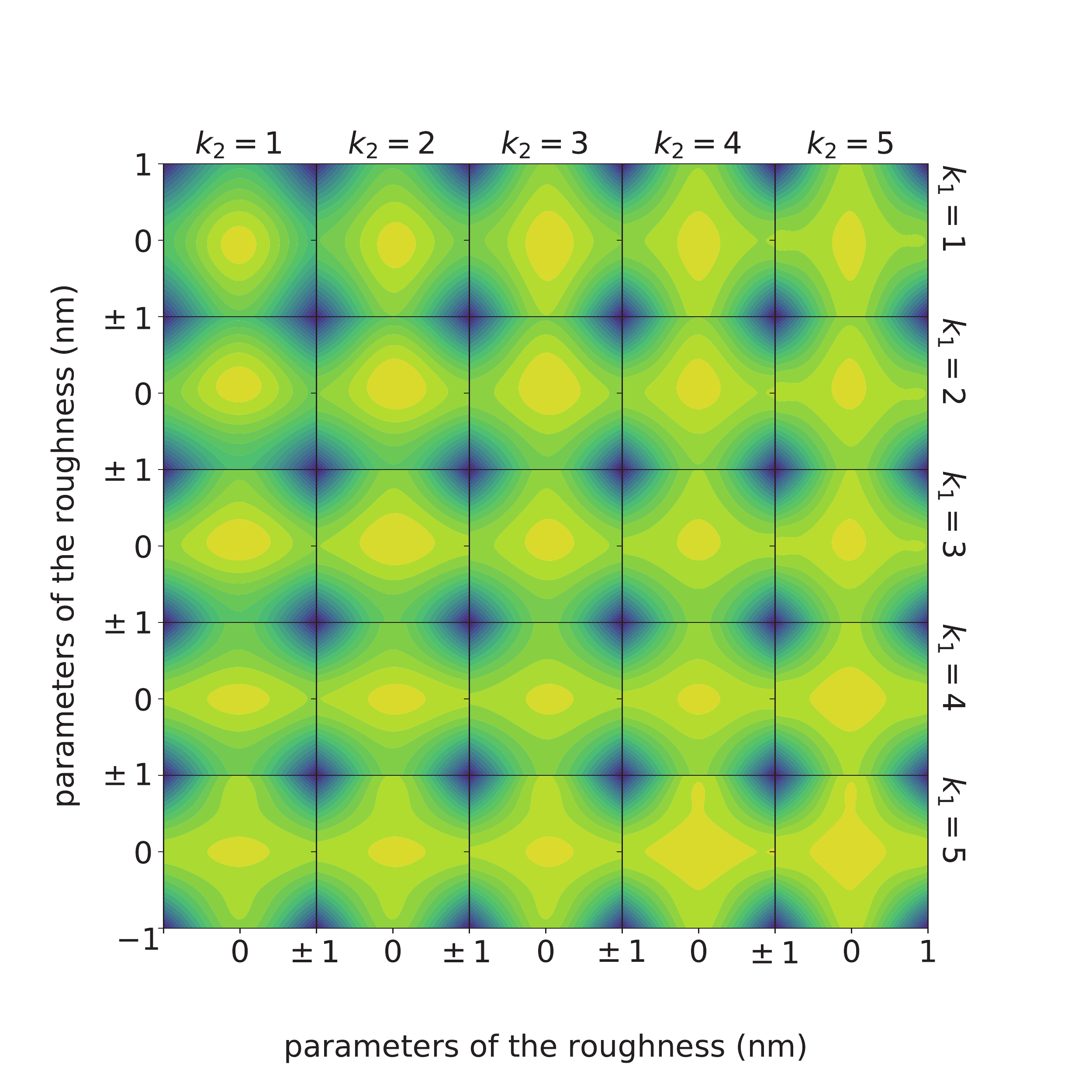}
	}
	\caption{Contour plots of exciton diffusion length on random variables in 3D when $\beta=-2, 0$, respectively.}\label{fig:lambdacontour}
\end{figure}

\textbf{\textit{Landscape exploration}}
Contour plots of the fitted EDL on random variables are given in Fig. \ref{fig:lambdacontour} when $\beta=-2, 0$, respectively. In each subfigure, EDL $\sigma$ is plotted as a function of $\theta_{k_1}(\omega_1)$ and $\theta_{k_2}(\omega_2)$, where $k_1,k_2=1,2,3,4,5$ and all the remaining random variables are set to be $0$. A direct comparison between $\beta=-2$ and $\beta=0$ in Fig. \ref{fig:lambdacontour} illustrates the directional (anisotropic) dependence of EDL on random variables, due to different decay rates of random variables in the surface roughness.
\par
\begin{figure}
\centering
\subfigure[$\theta(\omega_2)$ fixed]{
\includegraphics[width=0.45\textwidth]{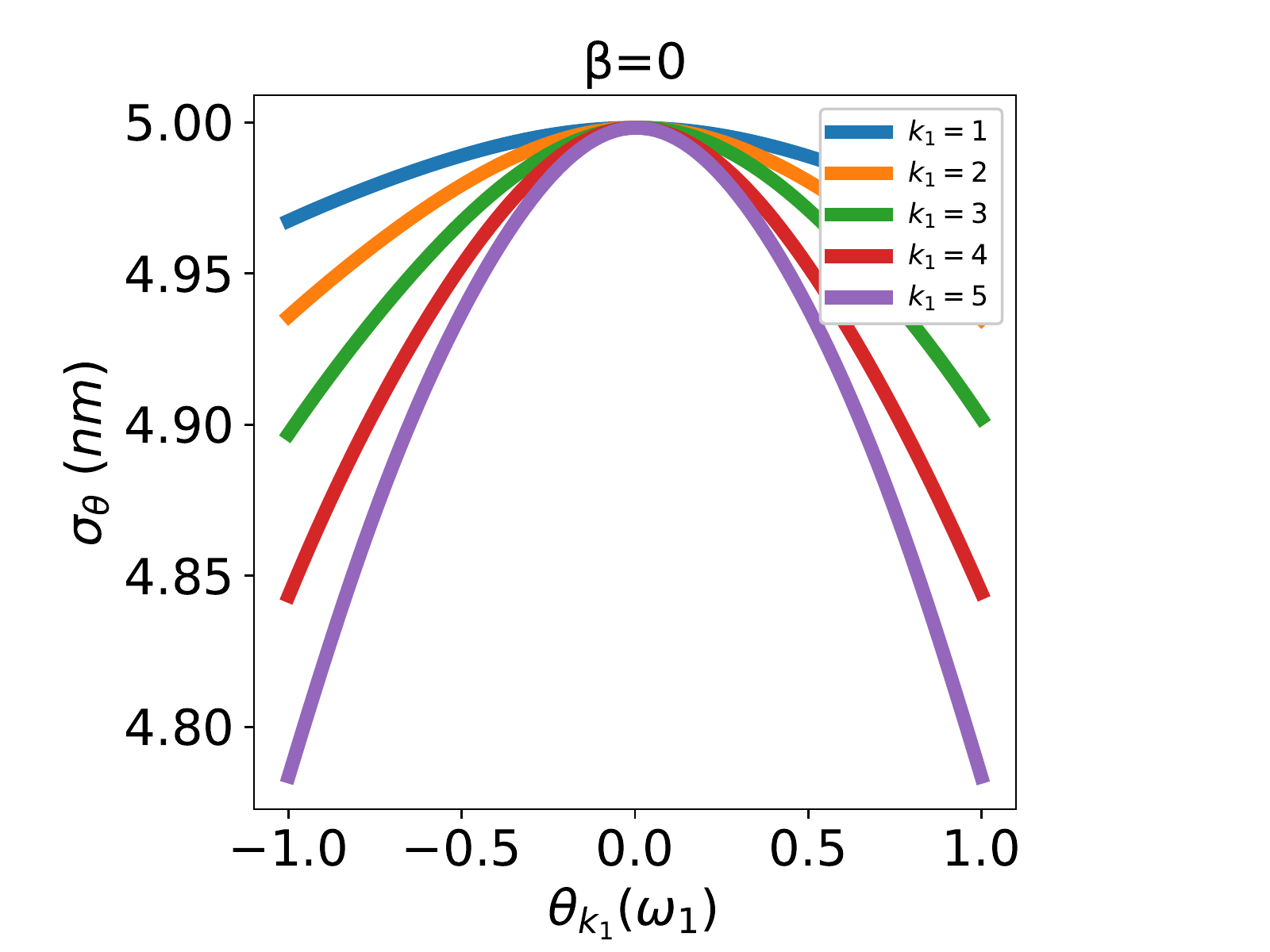}
}
\quad
\subfigure[$\theta(\omega_1)$ fixed]{
\includegraphics[width=0.45\textwidth]{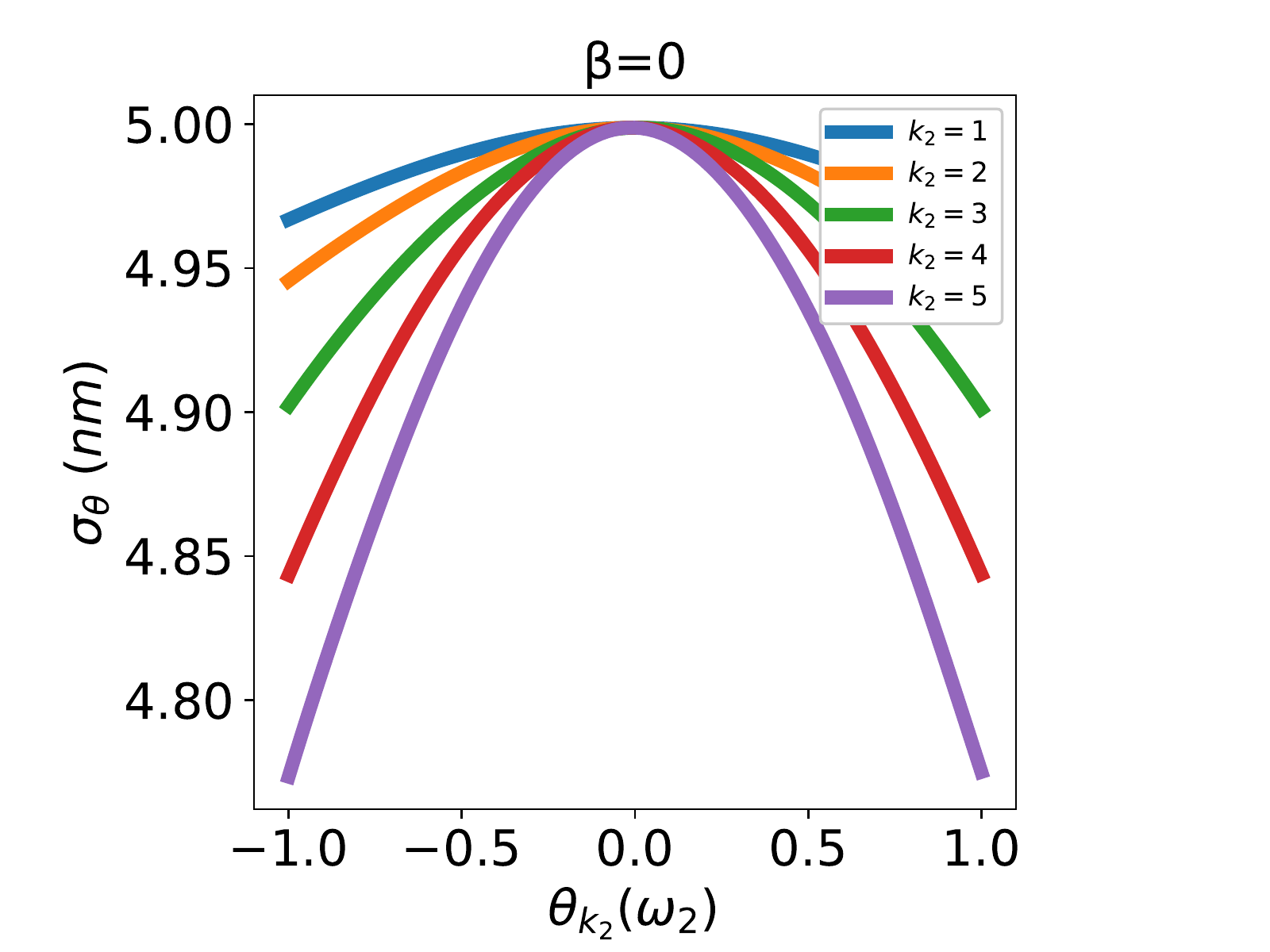}
}
\quad
\subfigure[$\theta(\omega_2)$ fixed]{
\includegraphics[width=0.45\textwidth]{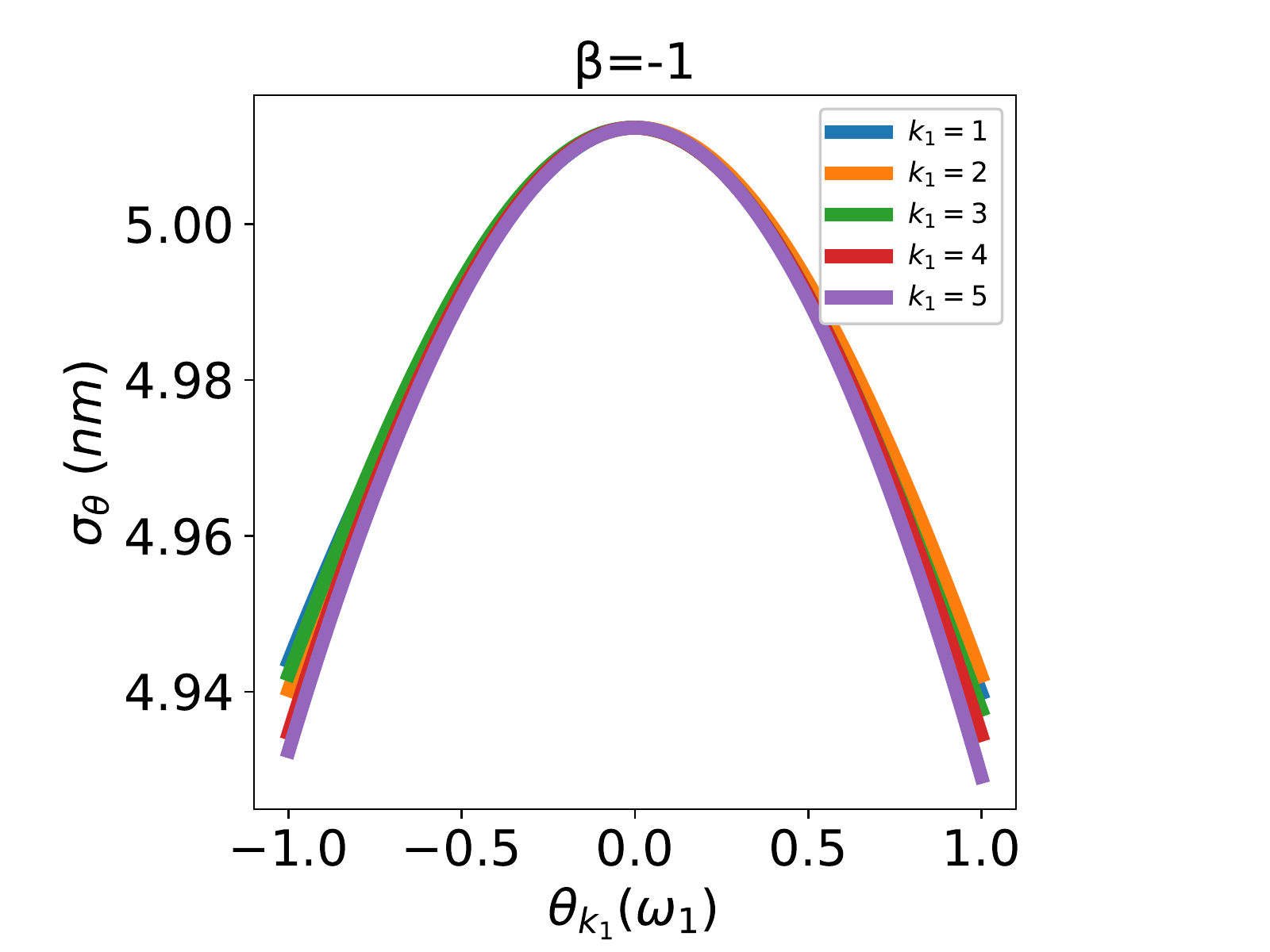}
}
\quad
\subfigure[$\theta(\omega_1)$ fixed]{
\includegraphics[width=0.45\textwidth]{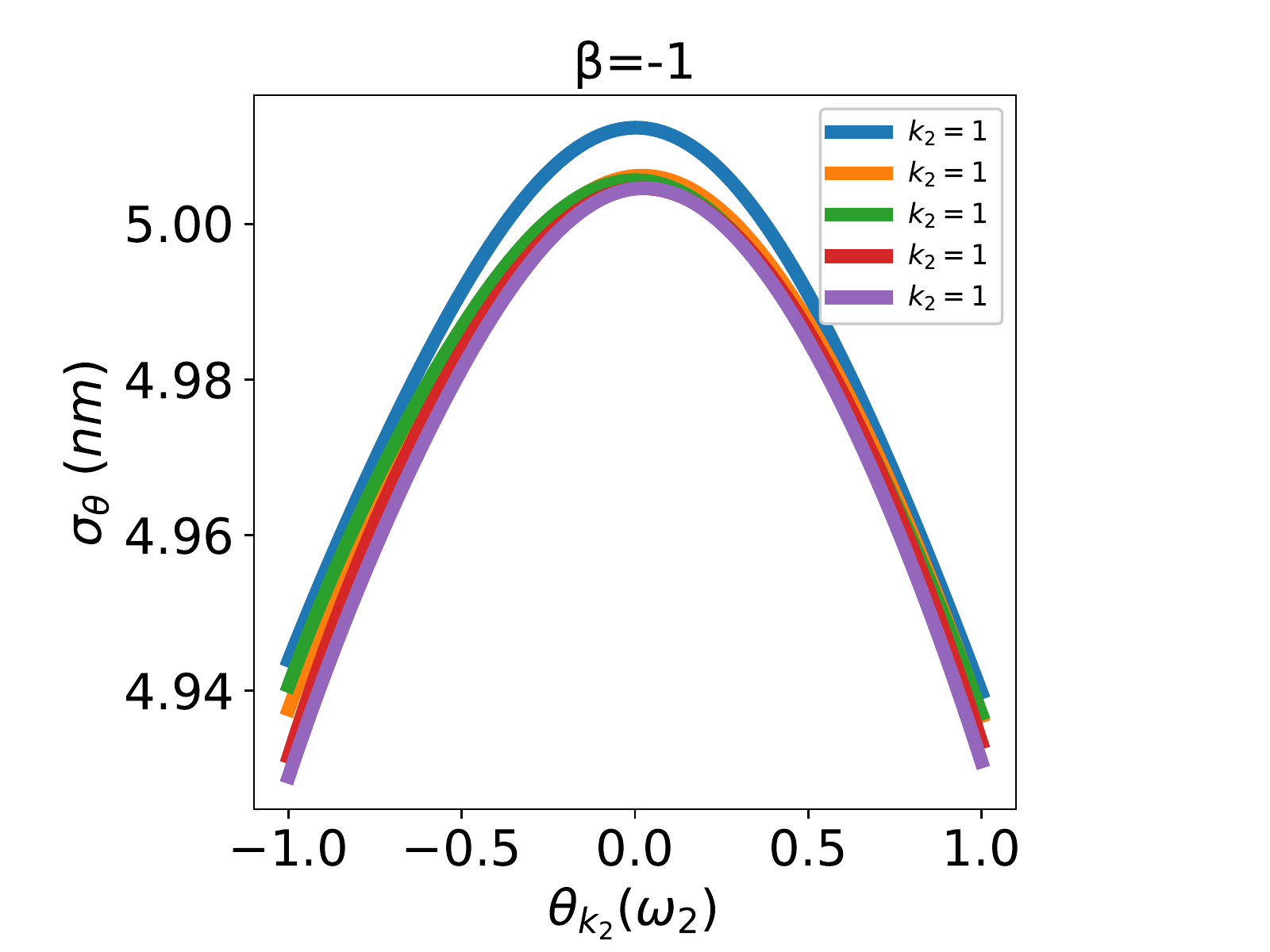}
}
\quad
\subfigure[$\theta(\omega_2)$ fixed]{
\includegraphics[width=0.45\textwidth]{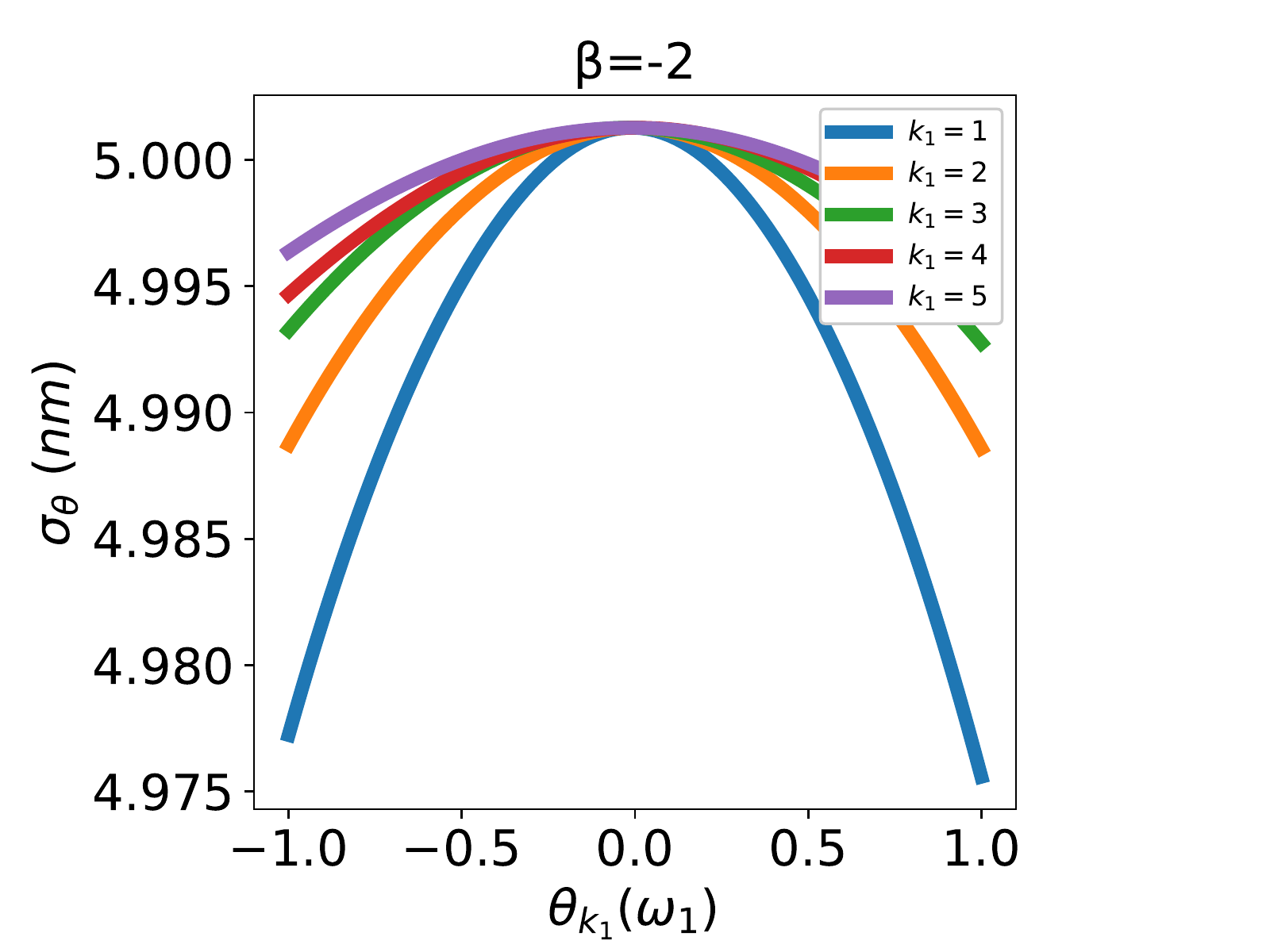}
}
\quad
\subfigure[$\theta(\omega_1)$ fixed]{
\includegraphics[width=0.45\textwidth]{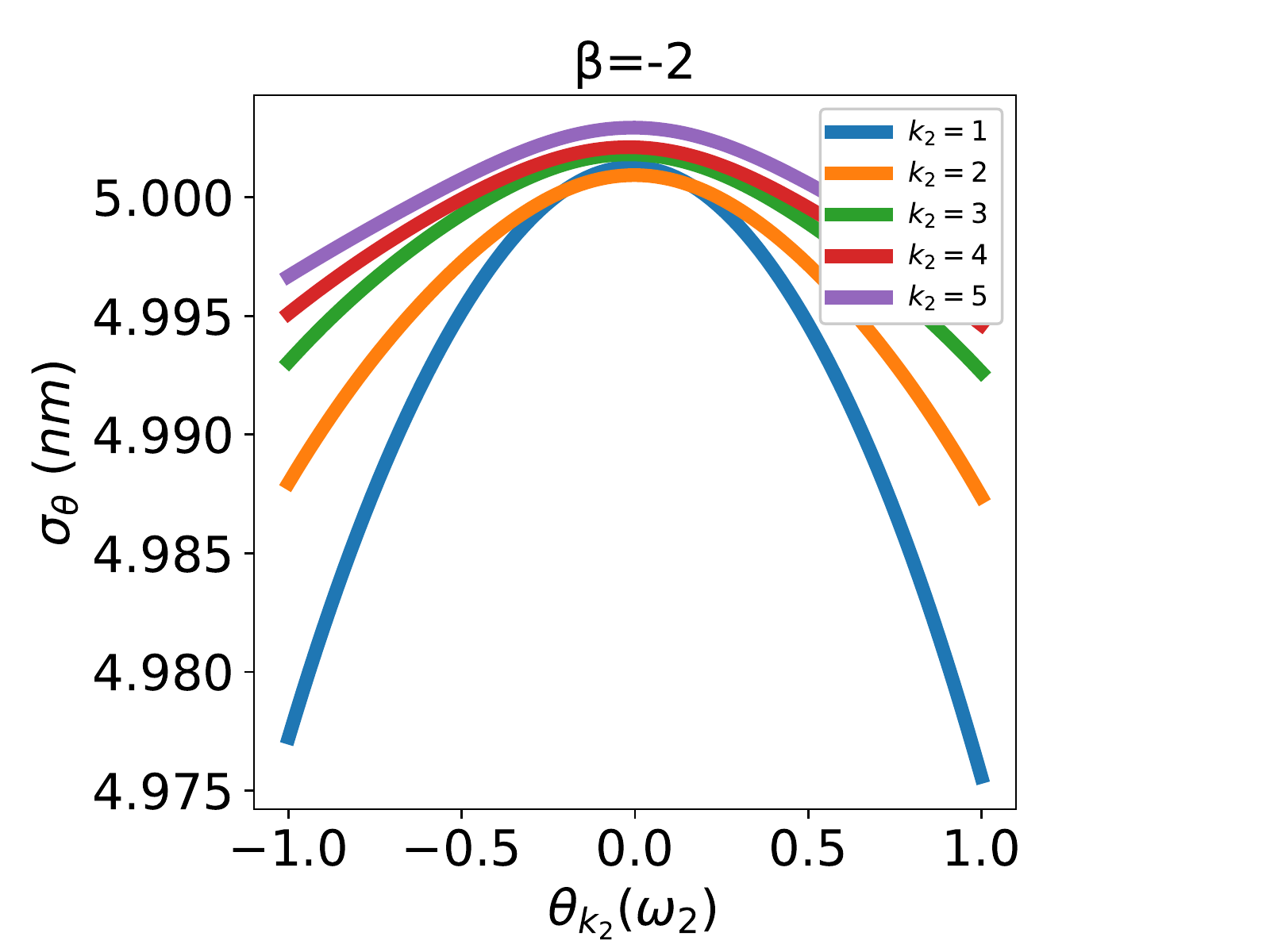}
}
\caption{Detailed dependence of exciton diffusion length on random variables. Top row: $\beta=-2$; Middle row: $\beta=-1$; Bottom row: $\beta=0$.
Left column: $\theta(\omega_2)=[1;0;0;0;0]$ is fixed and $\sigma$ is plotted as a function of $\theta_{k_1}(\omega_1)\in [-1,1]$, where $k_1=1,2,3,4,5$; Right column: $\theta(\omega_1)=[1;0;0;0;0]$ is fixed and $\sigma$ is plotted as a function of $\theta_{k_1}(\omega_1)\in [-1,1]$, where $k_1=1,2,3,4,5$.}
\label{fig:lambda0d1}
\end{figure}

\textbf{\textit{Mode dependence}}
Fig. \ref{fig:lambda0d1} provides a detailed demonstration of the dependence of EDL on random variables for $\beta=-2,-1,0$. For illustration, we keep $\theta(\omega_2)=[1;0;0;0;0]$ fixed in the left column and $\theta(\omega_1)=[1;0;0;0;0]$ fixed in the right column. One distinct difference between 3D and 2D is that the maximum EDL is approached in the absence of randomness in 3D, in contrast to the minimum EDL in 2D. The 3D result is reasonable since experimentally larger EDL is observed if the effect of surface roughness is minimized, while the 2D result is also of interest due to the unique dimensional dependence. When $\beta=-2$, the EDL is more sensitive to the lower-order modes (smaller $k$) and is less sensitive to the high-order modes (larger $k$). When $\beta=0$, the trend is completely opposite. This observation provides a detailed connection between surface roughness and EDL, which also sheds light on the experimental design. Given a surface roughness characterized by \eqref{eqn:3DInterface},  we have the value of $\beta$, from which we know which mode is of the most importance. Consequently, targeted experimental techniques can be applied to improve the opto-electronic performance.

%-----------------------------------------------------------------------------
\section{Conclusion}
\label{Sec:Conclusion}
In summary, we have developed a novel method based on quasi-Monte Carlo sampling and ResNet to approximate the exciton diffusion length in terms of surface roughness parametrized by a high-dimensional random field. This method extracts a function for exciton diffusion length over the entire parameter space. Rich information, such as landscape profile and mode dependence, can be extracted with unprecedented details. Useful information regarding the modeling error and the experimental design can be provided, which sheds lights on how to reduce the modeling error and how to design better experiments to improve opto-electronic properties of organics materials.

\section{Acknowledgments}
L. Lyu acknowledges the financial support of Undergraduate Training Program for Innovation and Entrepreneurship, Soochow University (Projection 201810285019Z). Z. Zhang acknowledges the financial support of Hong Kong RGC grants (Projects 27300616, 17300817, and 17300318) and  National Natural Science Foundation of China via grant 11601457, Seed Funding Programme for Basic Research (HKU), and Basic Research Programme of The Science, Technology and Innovation Commission of Shenzhen Municipality (JCYJ20180307151603959). J. Chen acknowledges the financial support by National Natural Science Foundation of China via grants 21602149, 11971021, and National Key R\&D Program of China (No. 2018YFB0204404). Part of the work was done when J. Chen was visiting Department of Mathematics, University of Hong Kong. J. Chen would like to thank its hospitality.

%%Vancouver style references.
\newpage
\bibliographystyle{plain}
\bibliography{refs}

\end{document}